%% file: Darboux-II.tex
\documentclass[11pt]{article} 
\usepackage{epsfig}
\begin{document}
\input posfig.tex
\input amssym.def 
\input amssym
\hfuzz=5.0pt
%
%
%
%
\def\vec#1{\mathchoice{\mbox{\boldmath$\displaystyle\bf#1$}}
{\mbox{\boldmath$\textstyle\bf#1$}}
{\mbox{\boldmath$\scriptstyle\bf#1$}}
{\mbox{\boldmath$\scriptscriptstyle\bf#1$}}}
\def\mbf#1{{\mathchoice {\hbox{$\rm\textstyle #1$}}
{\hbox{$\rm\textstyle #1$}} {\hbox{$\rm\scriptstyle #1$}}
{\hbox{$\rm\scriptscriptstyle #1$}}}}
\def\operatorname#1{{\mathchoice{\rm #1}{\rm #1}{\rm #1}{\rm #1}}}
\chardef\ii="10
\def\widehat{\mathaccent"0362 }
\def\widetilde{\mathaccent"0365 }
\def\vphi{\varphi}
\def\vrho{\varrho}
\def\vtheta{\vartheta}
\def\ih{{\i\over\hbar}}
\def\hi{\frac{\hbar}{\i}}
\def\CD{{\cal D}}
\def\CE{{\cal E}}
\def\CH{{\cal H}}
\def\CL{{\cal L}}
\def\CP{{\cal P}}
\def\CV{{\cal V}}
\def\half{{1\over2}}
\def\bhalf{\hbox{$\half$}}
\def\viert{{1\over4}}
\def\bviert{\hbox{$\viert$}}
\def\hhbox#1#2{\hbox{$\frac{#1}{#2}$}}
\def\dfrac#1#2{\frac{\displaystyle #1}{\displaystyle #2}}
\def\intT{\ih\int_0^\infty\d\,T\,e^{\i ET/\hbar}}
\def\pathint#1{\int\limits_{#1(t')=#1'}^{#1(t'')=#1''}\CD #1(t)}
\def\hbarm{{\dfrac{\hbar^2}{2m}}}
\def\hbarmq{{\dfrac{\hbar^2}{2mq}}}
\def\mzwei{\dfrac{m}{2}}
\def\overh{\dfrac1\hbar}
\def\ihbar{\dfrac\i\hbar}
\def\intt{\int_{t'}^{t''}}
\def\tn{\tilde n}
\def\pmb#1{\setbox0=\hbox{#1}
    \kern-.025em\copy0\kern-\wd0
    \kern.05em\copy0\kern-\wd0
    \kern-.025em\raise.0433em\box0}
\def\pathintG#1#2{\int\limits_{#1(t')=#1'}^{#1(t'')=#1''}\CD_{#2}#1(t)}
\def\limN{\lim_{N\to\infty}}
\def\Norm{\bigg({m\over2\pi\i\epsilon\hbar}\bigg)}
\def\hbaram{{\hbar^2\over8m}}
\def\bbbr{{\rm I\!R}}                                
\def\bbbn{{\rm I\!N}}                                
\def\bbbz{{\mathchoice {\hbox{$\sf\textstyle Z\kern-0.4em Z$}}
{\hbox{$\sf\textstyle Z\kern-0.4em Z$}}
{\hbox{$\sf\scriptstyle Z\kern-0.3em Z$}}
{\hbox{$\sf\scriptscriptstyle Z\kern-0.2em Z$}}}}    
\def\bbbc{{\mathchoice {\setbox0=\hbox{\rm C}\hbox{\hbox
to0pt{\kern0.4\wd0\vrule height0.9\ht0\hss}\box0}}
{\setbox0=\hbox{$\textstyle\hbox{\rm C}$}\hbox{\hbox
to0pt{\kern0.4\wd0\vrule height0.9\ht0\hss}\box0}}
{\setbox0=\hbox{$\scriptstyle\hbox{\rm C}$}\hbox{\hbox
to0pt{\kern0.4\wd0\vrule height0.9\ht0\hss}\box0}}
{\setbox0=\hbox{$\scriptscriptstyle\hbox{\rm C}$}\hbox{\hbox
to0pt{\kern0.4\wd0\vrule height0.9\ht0\hss}\box0}}}}
\def\CP{{\cal P}}
\def\CQ{{\cal Q}}
\def\Ai{\operatorname{Ai}} 
\def\Cl{\operatorname{Cl}} 
\def\SU{\operatorname{SU}} 
\def\dt{\d t}
\def\d{\operatorname{d}}
\def\e{\operatorname{e}}
\def\i{\operatorname{i}}
\def\sn{\operatorname{sn}}
\def\cn{\operatorname{cn}}
\def\max{\operatorname{max}}
\def\DI{D_{\,\rm I}}
\def\DII{D_{\,\rm II}}
\def\3dDII{D_{\,3d-\rm II}}
\def\DIII{D_{\,\rm III}}
\def\DIV{D_{\,\rm IV}}
\def\vphi{\varphi}
\def\tvphi{{\tilde\varphi}}
\def\tomega{{\tilde\omega}}
\def\ttau{{\tilde\tau}}
\def\hvphi{{\hat\varphi}}
\def\homega{{\hat\omega}}
\def\htau{{\hat\tau}}
\def\ps{\operatorname{ps}}
\def\Ps{\operatorname{Ps}}
\def\Si{\operatorname{Si}}
\def\energyldrei{\e^{-\i\hbar T(p^2+1)/2m}}
\def\ints{\int_0^{s''}}
\def\OO{\operatorname{O}}
\def\SO{\operatorname{SO}}
\def\fg{{\frak g}}
\def\fs{{\frak s}}
\def\fl{{\frak l}}
\def\gsl{\fg\fs\fl}
\def\operatorname#1{{\mathchoice{\rm #1}{\rm #1}{\rm #1}{\rm #1}}}
\def\bbbone{{\mathchoice {\rm 1\mskip-4mu l} {\rm 1\mskip-4mu l}
{\rm 1\mskip-4.5mu l} {\rm 1\mskip-5mu l}}}
\def\pathint#1{\int\limits_{#1(t')=#1'}^{#1(t'')=#1''}\CD #1(t)}
\def\pathints#1{\int\limits_{#1(0)=#1'}^{#1(s'')=#1''}\CD #1(s)}
 
\begin{titlepage}
\centerline{\normalsize DESY 06--113 \hfill ISSN 0418 - 9833}

\vskip.3in
\message{TITLE:}
\begin{center}
{\bf\Large Path Integral Approach for Superintegrable Potentials on Spaces
\\[3mm]
of Non-constant Curvature: I. Darboux Spaces $\DI$ and $\DII$.}
\end{center}
\message{Path Integral Approach for Superintegrable Potentials
on Spaces of Non-constant Curvature: I. Darboux Spaces DI and DII.}
\vskip.3in
\begin{center}
{\large Christian Grosche}
\vskip.1in
{\normalsize\em II.\,Institut f\"ur Theoretische Physik}
\vskip.05in
{\normalsize\em Universit\"at Hamburg, Luruper Chaussee 149}
\vskip.05in
{\normalsize\em 22761 Hamburg, Germany}
\end{center}
\vskip.2in
\begin{center}
{\large George S.\,Pogosyan}
\vskip.1in
{\normalsize\em Laboratory of Theoretical Physics}
\vskip.05in
{\normalsize\em Joint Institute for Nuclear Research (Dubna)}
\vskip.05in
{\normalsize\em 141980 Dubna, Moscow Region, Russia}
\vskip.05in
{\normalsize\em and}
\vskip.05in
{\normalsize\em Departamento de Matematicas}
\vskip.05in
{\normalsize\em CUCEI, Universidad de Guadalajara}
\vskip.05in
{\normalsize\em Guadalajara, Jalisco, Mexico}
\vskip.2in
{\large Alexei N.\,Sissakian}
\vskip.1in
{\normalsize\em Laboratory of Theoretical Physics}
\vskip.05in
{\normalsize\em Joint Institute for Nuclear Research (Dubna)}
\vskip.05in
{\normalsize\em 141980 Dubna, Moscow Region, Russia}
\end{center}
\normalsize
\vfill
\begin{center}
{\bf Abstract}
\end{center}
In this paper the Feynman path integral technique is applied for
superintegrable potentials on two-dimensional spaces of non-constant
curvature: these spaces are Darboux spaces $\DI$ and $\DII$, 
respectively. On $\DI$ there are three and on $\DII$ 
four such potentials, respectively.
We are able to evaluate the path integral in most of the
separating coordinate systems, leading to expressions for the Green functions,
the discrete and continuous wave-functions, and the discrete
energy-spectra. In some cases, however, the discrete spectrum cannot be stated
explicitly, because it is either determined by a transcendental
equation involving parabolic cylinder functions 
(Darboux space I), or by a higher order polynomial equation.
The solutions on $\DI$ in particular show that superintegrable systems
are not necessarily degenerate.
We can also show how the limiting cases of flat space (constant
curvature zero) and the two-dimensional hyperboloid (constant negative
curvature) emerge.

\end{titlepage}
 
 
\tableofcontents

\bigskip\bigskip

\setcounter{page}{1}%
\setcounter{equation}{0}%
\section{Introduction}%
\message{Introduction}%

\subsubsection*{General Overview and Recent Work}
\message{General Overview and Recent Work}
In the last years an enormous amount of work has been archived in 
solving path integrals in quantum mechanics exactly, and to the application 
of the path integral method in various branches of mathematical physics; 
many of this has been compiled in our publication \cite{GRSh}. 
In \cite{GROad} one of us have discussed path integral
representations of the free motion in two and three dimensions for Euclidean
space, Pseudo-Euclidean space, spheres and hyperboloids.
In these studies it was the goal to find all path integral representations
for the coordinate systems \cite{KAL,MILLf}--\cite{OLE} 
in which the Schr\"odinger equation
respectively the path integral allows separation of variables.
\cite{GRSh} was devoted to give a best to our knowledge list of up-to-date
explictly known path integral solutions. 

In the present work we extend our studies of superintegrable potentials to
spaces of non-constant curvature, 
i.e. Darboux spaces, by means of the path integral method.
In the following sections we discuss two Darboux spaces:
we set up the Lagrangian, the Hamiltonian, the quantum operator, and 
formulate and solve (if this is possible) the corresponding path integral.
We also discuss some of the limiting cases of the Darboux-spaces,
i.e. where we obtain a space of constant (zero or negative) curvature.
In the case of $\DI$ there is no limiting case, because we have no free
parameter in the metric to choose from.  

In a recent publication one of us \cite{GROas} 
has applied the path integral technique \cite{FH,GRSh,KLEo,SCHUHd}
to the quantum motion on two-dimensional spaces of non-constant curvature,
called Darboux spaces,  $\DI$--$\DIV$, respectively. 
These spaces have been introduced by Kalnins et 
al.~\cite{KalninsKMWinter,KalninsKWinter}. They can be embedded in 
three-dimensional spaces which can be either of Euclidean or Minkowskian type, 
respectively.
Then the Darboux spaces consist of surfaces, which are also called
surfaces of revolution \cite{DASYPS}. In two dimensions Darboux spaces of
non-constant curvature can be constructed as follows. One takes for instance
two-dimensional Euclidean space and takes for the metric a superintegrable
potential in its simplest form in radial coordinates. For the Coulomb
potential $1/r$ one obtains a metric $\propto r$, which gives the Darboux
space $\DI$, for the radial potential $b-a/r^2$ one obtains a metric $\propto
(b-a/r^2)$, i.e. the  Darboux space $\DII$, etc. The case of two-dimensions is
especially simple, because one obtains always a conformally flat space. This
method to construct new spaces was first discussed by Koenigs \cite{KOENIGS}.

\subsubsection*{Superintegrable Potentials}
\message{Superintegrable Potentials}
The intention of \cite{KalninsKMWinter,KalninsKWinter} was however, not only
to construct new spaces, and to study their properties, but another equally
important motivation was to find the corresponding superintegrable
potentials. The notion of superintegrable systems was introduced by 
Winternitz and co-workers in \cite{FMSUW,WSUF}, Wojciechowski \cite{WOJ}, 
and was developed further 
later on also by Evans \cite{EVA}. Superintegrable potentials have the 
property that one finds additional constants of motion: The simples case of
the case of the only conserved quantity is the energy gives usually a 
chaotic system \cite{GRSh}; in order that a physical system is just integrable
requires $d$ constants of motion, where $2d$ denotes the number of degrees 
of freedom. In two dimensions one
obtains in total three functional independent constants of motion and in three
dimensions one has four (minimal superintegrable) and five (maximal
superintegrable) functional independent constants of
motion. Well-known examples are the Coulomb potential with its Lenz--Runge
vector and the harmonic oscillator with its quadrupole moment.

Moreover, the existence of an additional conserved quantity in (maximally) 
superintegrable potentials leads in classical mechanics to the fact that the 
orbits of a particle in such a potential are closed: Kepler ellipses are stable
and do not ''rotate''. In quantum mechanics it follows that the spectrum is 
usually degenerate.
A perturbation of the pure Newtonian potential causes the Kepler ellipses 
to rotate (Mercury's or the Moon's perihelion rotation), 
and in quantum mechanics degeneration is lost, respectively.

Another feature of superintegrable potentials is that the corresponding
equations in classical and quantum mechanics separate in more than one
coordinate system. (However, whereas from the separability in more that one 
coordinate system the superintegrability and the existence of additional
constants of motion follows, a system with additional constants of motion 
may not be easily separable.) 
It turns out that the Coulomb potential in three dimensions separates in
spherical, conical, parabolic and prolate-spheroidal coordinates
\cite{MPST-A}.
Even the relativistic Dirac-Coulomb possesses some of this symmetry by the
conservation of the Johnson--Lippmann operator which reduces in the
non-relativistic limit to the Lenz--Runge vector \cite{KHKH}.

In previous publications \cite{GROPOa}--\cite{GROPOd} we have studies
superintegrable potentials in two and three dimensions in Euclidean space, on
spheres and on hyperboloids. We restricted ourselves to real spaces and
omitted their corresponding complex extensions 
\cite{KKPM,KMP4,KMP1,KMP2}. Let us also note that by integrating
out ignorable coordinates (i.e. variables which have plane waves, respectively
circular waves as solutions 
of the Schr\"odinger equation) one can obtain from a higher dimensional
more complicated space interacting systems on spaces with constant curvature:
the interaction has the form of a superintegrable  potential.
One example is the hermitian hyperbolic space \cite{BKW,GROar} where one can
find superintegrable  potentials on the hyperboloid \cite{KMP5}. 
The connection with superintegrability and the polynomial solutions was 
studied e.g.~in \cite{KMPa}, in the connections with contractions of Lie
algebras e.g.~in \cite{IPSW,KMPb,PGW}, where the various limiting cases from
spaces of positive or negative constant curvature to zero curvature was
investigated.

In this first paper on super integrable potentials on Darboux spaces we
discuss only the Darboux spaces $\DI$ and $\DII$.
The superintegrable potentials on the other two Darboux spaces $\DIII$ and
$\DIV$ will be discussed in a forthcoming  publication.

The paper is organized as follows: In the next Section we treat the 
superintegrable potentials on Darboux space $\DI$. There are three of them,
the third consisting of a constant divided by the metric term which makes the
potential almost trivial. 
The common features of the first two potentials is that
the energy eigenvalues are determined by a transcendental equation involving
parabolic cylinder functions. For the third (trivial) potential no bound states can be
found.

In the third section the superintegrable potentials on $\DII$ are discussed.
There are three non-trivial potentials and one trivial.
For the first potential we obtain a quadratic equation for the energy-levels,
and they show an oscillator-like behavior. An exact solution can be found
only in the $(u,v)$-system. This is very similar to the Holt
potential in two-dimensional Euclidean space.

The second superintegrable potential on $\DII$ is exactly solvable in two
coordinate systems. Here, we also find a quadratic equation for the energy
levels. $V_2$ is similar to the singular oscillator in two-dimensional
Euclidean space. 
  
The third superintegrable potential has a relation to the Coulomb potential 
in two-dimensional Euclidean space. The energy levels are determined by an
equation of eight order in $E$ which cannot be solved in general. For a
special case, however, we find a Coulomb-like behavior of the energy-levels.

The fourth potential is a constant times the metric term, and is therefore
trivial. As for $\DI$ this potential is included for completeness.

The fourth Section contains a discussion of our results and an outlook for the 
remaining two Darboux spaces $\DIII$ and $\DIV$.

\subsubsection*{Introducing Darboux Spaces}
\message{Introducing Darboux Spaces}
Kalnins et al. \cite{KalninsKMWinter,KalninsKWinter} denoted four types of 
two-dimensional spaces of non-constant curvature, labeled by
$\DI$--$\DIV$, which are called Darboux spaces \cite{KOENIGS}. 
In terms of the infinitesimal distance they are described~by
(the coordinates $(u,v)$ will be called the $(u,v)$-system; 
the $(x,y)$-system in turn can be called light-cone coordinates):
\begin{eqnarray}
({\rm I})  \qquad  \d s^2&=& (x+y)\d x\d y
\nonumber\\
&=&2u(\d u^2+ \d v^2)\enspace,\qquad (x=u+\i v,y=u-\i v)\enspace,
\label{DarbouxI}
\\[2mm]
({\rm II}) \qquad \d s^2&=& \bigg(\frac{a}{(x-y)^2}+b\bigg) \d x\d y
\nonumber\\
 &=&\frac{bu^2-a}{u^2}(\d u^2+\d v^2)\enspace,\qquad
\Big(x=\bhalf(v+\i u), y=\bhalf(v-\i u)\Big)\enspace,
\label{DarbouxII}
\\[2mm]
({\rm III})\qquad \d s^2&=& \big(a\,\e^{-(x+y)/2}+b\,\e^{-x-y}\big)
\d x\d y
\nonumber\\
&=&\e^{-2u}(b+a\,\e^u)(\d u^2+\d v^2)\enspace,
\qquad (x=u-\i v,y=u+\i v)\enspace,
\label{DarbouxIII}
\\[2mm]
({\rm IV}) \qquad \d s^2&=& -\frac{a\big(\e^{(x-y)/2}+\e^{(y-x)/2}\big)+b}
{\big(\e^{(x-y)/2}-\e^{(y-x)/2}\big)^2}\d x\d y
\nonumber\\
&=&\left(\frac{a_+}{\sin^2u}+\frac{a_-}{\cos^2u}\right)(\d u^2+\d v^2)\,
\qquad (x=u+\i v,y=u-\i v)\enspace.
\label{DarbouxIV}
\end{eqnarray}
$a$ and $b$ are additional (real) parameters ($a_\pm=(a\pm 2b)/4$).
Kalnins et al. \cite{KalninsKMWinter,KalninsKWinter} studied not only the
solution of the free motion, but also emphasized on the superintegrable
systems in theses spaces. They found appropriate
coordinate systems, and we will consider all of them. In the majority
of the cases we will be able to find a solution, however in some cases
this will not be possible due to a quartic anharmonicity of the
problem in question. 

\goodbreak\noindent%

\setcounter{equation}{0}%
\section{Superintegrable Potentials on Darboux Space $\DI$}%
\message{Superintegrable Potentials on Darboux Space D_I}%
We start with Darboux Space $\DI$ and
consider the following coordinate systems
\begin{eqnarray}
\hbox{($(u,v)$-System:)}&&
x=u+\i v\enspace,\quad y=u-\i v\enspace,\qquad\,\,
(u\geq a)\,,
\\
\hbox{(Rotated $(r,q)$-Coordinates:)}&&
u=r\cos\vtheta+q\sin\vtheta\enspace, \nonumber\\  &&
v=-r\sin\vtheta +q\cos\vtheta,\qquad\qquad\,\,\,\, (\vtheta\in[0,\pi]),
\\
\hbox{(Displaced parabolic:)}&&
u=\half(\xi^2-\eta^2)+c\enspace,\,\,
v=\xi\eta\enspace,\quad (\xi\in\bbbr,\eta>0,c>0)\,.\qquad
\end{eqnarray}
The infinitesimal distance, i.e., the metric is given by
\begin{eqnarray}
\d s^2&=&2u(\d u^2+ \d v^2)\enspace,\\
\hbox{(Rotated $(r,q)$-Coordinates:)}
&=&2(r\cos\vtheta+q\sin\vtheta)(\d r^2+ \d q^2)\enspace,\\
\hbox{(Displaced parabolic:)}
&=&(\xi^2-\eta^2+2c)(\xi^2+\eta^2)(\d\xi^2+ \d\eta^2)\enspace.
\qquad\qquad\qquad\qquad
\end{eqnarray}
The Gaussian curvature in a space with metric
$\d s^2=g(u,v)(\d u^2+\d v^2)$ is given by ($g=\det g(u,v)$)
\begin{equation}
G=-\frac{1}{2g}\bigg(\frac{\partial^2}{\partial u^2}
+\frac{\partial^2}{\partial v^2}\bigg)\ln g\enspace.
 \label{Gaussian-curvature}
\end{equation}
Equation (\ref{Gaussian-curvature}) will be used to discuss shortly the
curvature properties of the Darboux spaces, including their limiting cases of
constant curvature.

We find e.g. in the $(u,v)$-system for the Gaussian curvature
\begin{equation}
G=\frac{1}{u^4}\enspace.
\end{equation}

There is no further parameter in the metric, therefore this space is
of non-constant curvature throughout for all $u>a$ with $a$ some
real constant $a>0$. However, $\DI$ can be embedded in a three-dimensional
Euclidean space. It can then be visualized as an infinite surface (similar 
to one sheet of a double-sheeted hyberboloid) with a circular hole at the 
bottom \footnote{E.Kalnins, private communication.}.
The constant $a$ maybe taken as $a=\half$.
In order to set up the path integral formulation we follow our
canonical procedure as presented in \cite{GRSh}. The free Lagrangian and
Hamiltonian are given by, respectively:
\begin{equation}
\CL(u,\dot u, v,\dot v)=mu(\dot u^2+\dot v^2)-V(u,v), \quad
\CH(u,p_u,v,p_v)=\frac{1}{4mu}(p_u^2+p_v^2)+V(u,v)\enspace,
\end{equation}
and we must require $u>a$ for some $a>0$, and $\vphi\in[0,2\pi]$ can be
considered as a cyclic variable \cite{KalninsKWinter}.
The canonical momenta are
\begin{equation}
p_u=\hi\bigg(\frac{\partial}{\partial u}+\frac{1}{2u}\bigg),
\qquad 
p_v=\hi\frac{\partial}{\partial v}\enspace,
\end{equation}
and for the quantum Hamiltonian we find
\begin{eqnarray}
H=-\frac{\hbar^2}{2m}\frac{1}{2u}
\bigg(\frac{\partial^2}{\partial u^2}+\frac{\partial^2}{\partial
  v^2}\bigg)+V(u,v)
=\frac{1}{2m}\frac{1}{\sqrt{2u}}(p_u^2+p_v^2)\frac{1}{\sqrt{2u}}
+V(u,v)\enspace.
\end{eqnarray}
We formulate the path integral (ignoring the half-space
constraint for the time being):
\begin{eqnarray}
&&K(u'',u',v'',v';T)
\nonumber\\   &&=
\lim_{N\to\infty}\bigg(\frac{m}{2\pi\i\epsilon\hbar}\bigg)^N
\prod_{j=1}^{N-1}\int 2u_j\d u_j\d v_j
\exp\left\{\ih \sum_{j=1}^N\bigg[m
     \widehat{u_j}(\Delta^2u_j+\Delta^2v_j)-V(u_j,v_j)\bigg]\right\}\qquad
\\
&&=\pathint{u}\pathint{v}2u\exp\left\{\ih 
    \int_0^T\bigg[mu(\dot u^2+\dot v^2)-V(u,v)\bigg]\dt\right\}
\label{PathIntegralDI}
\end{eqnarray}
$u_j=u(t_j)$, $\Delta u_j=u_j-u_{j-1}$, $\epsilon=T/N$, 
$\widehat{u_j}=\sqrt{u_{j-1}u_j}$). 
We have displayed the path integral in our product-lattice definition,
which will be used throughout this paper \cite{GRSh}. Due to this 
lattice definition of the path integral, we have no additional 
$\hbar^2$-potential because the dimension of the space of 
non-constant curvature equals $2$, c.f.~\cite{GRSh}. 

\begin{table}[h!]
\caption{Constants of Motion in space $\DI$}
\label{ConstantMotionDI}
\begin{eqnarray}\begin{array}{l}\vbox{\small\offinterlineskip
\halign{&\vrule#&$\strut\ \hfil\hbox{#}\hfill\ $\cr
\noalign{\hrule}
height2pt&\omit&&\omit&&\omit&&\omit&&\omit&\cr
&Metric &&Constants of motion  &&Coordinate system &\cr
height2pt&\omit&&\omit&&\omit&&\omit&&\omit&\cr
\noalign{\hrule}\noalign{\hrule}
height2pt&\omit&&\omit&&\omit&&\omit&&\omit&\cr
&$2u(\d u^2+ \d v^2)$    &&$K^2$    &&$(u,v)$-System         &\cr
height2pt&\omit&&\omit&&\omit&&\omit&&\omit&\cr
\noalign{\hrule}
height2pt&\omit&&\omit&&\omit&&\omit&&\omit&\cr
&$2(r\cos\vtheta+q\sin\vtheta)(\d r^2+ \d q^2)$         
                         &&$X_1$    &&$(r,q)$-System    &\cr
height2pt&\omit&&\omit&&\omit&&\omit&&\omit&\cr
\noalign{\hrule}
height2pt&\omit&&\omit&&\omit&&\omit&&\omit&\cr
&$(\xi^2-\eta^2+2c)(\xi^2+\eta^2)(\d\xi^2+ \d\eta^2)$         
                         &&$X_2$    &&Parabolic         &\cr
height2pt&\omit&&\omit&&\omit&&\omit&&\omit&\cr
\noalign{\hrule}}}\end{array}\nonumber\end{eqnarray}
\end{table}

According to \cite{KalninsKMWinter,KalninsKWinter} we introduce the
following three integrals of motion in $\DI$. They are
\begin{equation}
\left.\begin{array}{rl}
K&=p_v\\
X_1&=p_up_v-\dfrac{v}{2u}(p_u^2+p_v^2)\\
X_2&=p_v(vp_u-up_v)-\dfrac{v^2}{4u}(p_u^2+p_v^2)\enspace.
\end{array}\qquad\right\}
\end{equation}
They satisfy the relation
\begin{equation}
4\tilde\CH_0 X_2+X_1^2+K^4=0\enspace.
\end{equation}
(Let us note that by $\tilde\CH_0$ the classical Hamiltonian without
the $1/2m$-factor is meant. Keeping this factor is no problem, however, 
in the present form the algebra has a simpler showing).
These operators satisfy the Poisson algebra relations
\begin{equation}
\{K,X_1\}=2\tilde\CH_0\enspace,\qquad
\{K,X_2\}=-X_1\enspace,\qquad
\{X_1,X_2\}=2K^3\enspace.
\end{equation}
The quantum analogues are given by
\begin{equation}
\left.\begin{array}{rl}
\widehat K&=\partial_v\\
\widehat X_1&=\partial_u\partial_v
   -\dfrac{v}{2u}(\partial_u^2+\partial_v^2)\\[3mm]
\widehat X_2&=\bhalf\{\partial_v,v\partial_u-u\partial_v\}
-\dfrac{v^2}{4u}(\partial_u^2+\partial_v^2)\enspace,
\end{array}\qquad\right\}
\end{equation}
where $\{\cdot,\cdot\}$ is the anti-commutator.
These operators satisfy the commutation relations
\begin{equation}
[\widehat K,\widehat X_1]=-2\widehat H_0\enspace,\qquad
[\widehat K,\widehat X_2]=X_1\enspace,\qquad
[\widehat X_1,\widehat X_2]=2{\widehat K}^3\enspace,
\end{equation}
with the operator relation
\begin{equation}
4\widehat H_0 \widehat X_2+\widehat X_1^2
+\widehat K^4=0\enspace.
\end{equation}
The operators $K,X_1,X_2$ can be used to characterize the separating
coordinate systems on $\DI$, as indicated in Table \ref{ConstantMotionDI}.

Let us note again that we do omit here factors of $\i$, $\hbar$ and $1/2m$
for the sake of simplicity. $H_0$ therefore is the quantum Hamiltonian
without the usual $-\hbar^2/2m$. However, in the tables with the
constants of motion, these factors are meant to be included.
In the remaining Darboux spaces this notation as long as the algebra 
is concerned will be for the sake of simplicity in the same way.

For the operators which characterize separation of variables in the
$(r,q)$-systems and parabolic coordinates, respectively, we introduce 
\begin{eqnarray}
\Lambda_1 &=&  \frac{1}{q\sin\theta + r\cos\theta}
\left(q\sin\theta \frac{\partial^2}{\partial r^2}
-  r\cos\theta \frac{\partial^2}{\partial q^2}\right)
\nonumber\\ &=&
 - \sin2\theta X_1 - \cos2\theta K^2\enspace,
\\
\Lambda_2 &=&  
\frac{1}{\xi^4 - \eta^4}\left(\eta^4 \frac{\partial^2}{\partial \xi^2}
+ \xi^4 \frac{\partial^2}{\partial \eta^2}\right)
+ \frac{4c \xi^2 \eta^2}{\xi^2-\eta^2}
\nonumber\\ &=&
-  \frac{\partial}{\partial u} - 2v  \frac{\partial}{\partial u}
\frac{\partial}{\partial v}
+ 2(u-c) \frac{\partial^2}{\partial v^2} + \frac{v^2}{2(u-c)}
\left( \frac{\partial^2}{\partial u^2} +  
\frac{\partial^2}{\partial v^2}\right)
+ \frac{2c v^2}{(u-c)}\enspace.
\end{eqnarray}
These two operators describe the general case.
Special cases for $\Lambda_1$ are:
\begin{itemize}
\item $\theta = \pi/4$, we have $\Lambda_1 = - X_1$ (symmetric case),
\item $\theta = \pi/2$, we have $\Lambda_1 =  K^2$,
\end{itemize}
and for $c=0$ we have $\Lambda_2 = - 2X_2$.

We now consider the following potentials on $\DI$
(following \cite{KalninsKWinter}, an additional  fourth potential 
is according to \cite{DASYPS}):
\begin{eqnarray}
V_1(u,v)&=&\frac{1}{2u}\Bigg[\frac{m}{2}\omega^2(4u^2+v^2)+\kappa+
\frac{\lambda^2-\viert}{2mv^2}\Bigg]\enspace,
\label{V1-D1}
\\   
V_2(u,v)&=&\frac{1}{2u}\bigg[\frac{m}{2}\omega^2(u^2+v^2)+\kappa_1+
\kappa_2v\bigg]\enspace,
\label{V2-D1}
\\
V_3(u,v)&=&\frac{1}{2u}\frac{\hbar^2v_0^2}{2m}\enspace,
\label{V3-D1}
\\
V_4(u,v)&=&\frac{1}{2u}\Bigg[
\frac{a_0}{\sqrt{u-\i v}}+a_1+a_2u+a_3\frac{4u-2\i v}{\sqrt{u-\i v}}
\Bigg]\enspace.
\label{V4-D1}
\end{eqnarray}

\begin{table}[t!]
\caption{Separation of variables for the
  superintegrable potentials on $\DI$}
\label{PotentialsDI}
\begin{eqnarray}\begin{array}{l}\vbox{\small\offinterlineskip
\halign{&\vrule#&$\strut\ \hfil\hbox{#}\hfill\ $\cr
\noalign{\hrule}
height2pt&\omit&&\omit&&\omit&&\omit&&\omit&\cr
&Potential&&Constants of motion  &&Separating coordinate system &\cr
height2pt&\omit&&\omit&&\omit&&\omit&&\omit&\cr
\noalign{\hrule}\noalign{\hrule}
height2pt&\omit&&\omit&&\omit&&\omit&&\omit&\cr
&$V_1$    &&$R_1=X_2-\frac{m}{2}\omega^2\frac{v^4}{4u}
                 -\frac{\kappa_2}{2}\frac{v^2}{u}
                 -\frac{\hbar^2}{4m}(\lambda^2-\viert)\frac{4u^2+v^2}{uv^2}$
                                   &&$\underline{\hbox{$(u,v)$-System}}$ &\cr
&         &&$R_2=K^2+\frac{m}{2}\omega^2v^2
                +\frac{\hbar^2}{m}\frac{\lambda^2-\viert}{v^2}$
                                   &&Parabolic                           &\cr
height2pt&\omit&&\omit&&\omit&&\omit&&\omit&\cr
\noalign{\hrule}
height2pt&\omit&&\omit&&\omit&&\omit&&\omit&\cr
&$V_2$    &&$R_1=X_1-\frac{\kappa_1v}{u}
                  +\frac{\kappa_2(u^2-v^2)}{u}
                  +\frac{m}{2}\omega^2\frac{v(u^2-v^2)}{u}$
                                    &&$\underline{\hbox{$(u,v)$-System}}$ &\cr
&         &&$R_2=K^2+2\kappa_2v+m\omega^2v^2$
                                    &&$\underline{\hbox{$(r,q)$-System}}$ &\cr
height2pt&\omit&&\omit&&\omit&&\omit&&\omit&\cr
\noalign{\hrule}
height2pt&\omit&&\omit&&\omit&&\omit&&\omit&\cr
&$V_3$   &&$R_1=X_1-\frac{\hbar^2v_0^2}{2m}\frac{v}{u}$
                                    &&$\underline{\hbox{$(u,v)$-System}}$ &\cr
&        &&$R_2=X_2-\frac{\hbar^2v_0^2}{4m}\frac{v^2}{u}$
                                    &&$\underline{\hbox{$(r,q)$-System}}$ &\cr
&        &&$R_3=K$                  &&$\underline{\hbox{Parabolic}}$      &\cr
 height2pt&\omit&&\omit&&\omit&&\omit&&\omit&\cr
\noalign{\hrule}}}\end{array}\nonumber\end{eqnarray}
\end{table}

In Table \ref{PotentialsDI} we have summarized some properties of
three of these potentials. Actually, $V_3$ can be considered as a special
case either of $V_1$ or $V_2$, respectively. The fourth potential
separates for instance in parabolic coordinates ($c=0$), and then has the 
(complex) form
\begin{equation}
V_4 (\xi, \eta) = \frac{1}{\xi^4-\eta^4} \left[ \sqrt{2}a_0 (\xi+i\eta)
+ a_1 (\xi^2+\eta^2) + \frac{a_2}{2} (\xi^4-\eta^4)
+ 2^{3/2}a_3 (\xi^3-i\eta^3)\right]\enspace.
\end{equation}
However, this is not tractable and we will not discuss this potential
any further.

\subsection{The Superintegrable Potential $V_1$ on $\DI$.}
\message{The Superintegrable Potential V_1 on D_I.}
We start with the potential $V_1$ in $\DI$. $V_1$ is separable in the 
$(u,v)$-system and in parabolic coordinates. However, only in the 
$(u,v)$-system a closed solution can be found. 
We state for  $V_1$ in the respective coordinate systems
\begin{eqnarray}
V_1(u,v)&=&\frac{1}{2u}\Bigg[\frac{m}{2}\omega^2(4u^2+v^2)+\kappa+
\frac{\lambda^2-\viert}{2mv^2}\Bigg]\enspace,
     \\ &=&
\frac{1}{2u(\xi^2+\eta^2)}\Bigg[\frac{m}{2}\omega^2(\xi^6+\eta^6)
+2m\omega^2(\xi^4-\eta^4)
\nonumber\\   &&\qquad\qquad\qquad\qquad
+(2m\omega^2c^2+\kappa)(\xi^2+\eta^2)
+\hbar^2\frac{\lambda^2}{2m}\bigg(\frac{1}{\xi^2}+\frac{1}{\eta^2}\bigg)
\Bigg]\enspace.\qquad  
\label{sextic-potential}
\end{eqnarray}
The separation procedure in the space-time transformation gives additional
terms according to
 $-E[(\xi^4-\eta^4)+2c(\xi^2+\eta^2)]$ in the respective Lagrangian.
Although symmetric in $\xi$ and $\eta$ the involvement of quartic and
sextic terms make any further evaluation impossible in parabolic coordinates. 

The same observations are valid in the case of a Coulomb-like potential 
on $\DI$, which can be put into the form 
(including already the proper energy term)
\begin{equation}
V_E(u,v)=-\frac{1}{u}\frac{\alpha}{\sqrt{u^2+v^2}}+E\enspace.
\end{equation}
which yields after a space-time transformation, with 
unshifted ($c=0$) parabolic coordinates
\begin{equation}
V_E(u,v)\rightarrow -2\alpha(\xi^2-\eta^2)+E(\xi^4-\eta^4)\enspace,
\end{equation}
and is not tractable either. In particular, the metric term $2u$ spoils
any further investigation. There exist some
attempts in the literature to treat such potential systems, and these
studies go with the name ``quasi-exactly solvable potentials'' in the sense
of Turbiner \cite{TURB} and Ushveridze \cite{USH}. 
In fact, sextic oscillators with a centrifugal barrier 
and quartic hyperbolic and trigonometric can be considered, and they are
very similar in their structure as for instance in (\ref{sextic-potential}).
One can find particular solutions, provided the parameters in the
quasi-exactly solvable potentials fulfill special conditions. Furthermore, 
well-defined expressions for the wave-functions and for the energy-spectrum 
can indeed be found if only quadratic, sextic, and a  particular centrifugal 
term are present. The wave-functions then have the form of $\Psi(x)\propto 
P(x^4)\times \e^{-\alpha x^4}$,  with a polynomial $P$.  However, 
quasi-exactly solvable potentials have the feature that only
a {\it finite} number of bound states can be calculated
(usually the ground state and some excited states).
Another important observation is due to \cite{KMP6,LETVIN}:
The authors found quasi-exactly solvable potentials that emerge from
dimensional reduction from two- and three-dimensional complex homogeneous
spaces. The sextic potential in the Hamiltonian (\ref{sextic-potential}) is 
exactly of that type. 

This observation now opens an interpretation of two-dimensional systems
with higher anharmonic terms. Let us assume that we have a two-dimensional
superintegrable potential system. This system has additional constants of
motion, respectively observables, and there are in total three of them
(including the energy). Let us assume further that we choose an example which
is separable in at least two coordinate systems, say in Cartesian and
parabolic coordinates (i.e., a system which is similar to the one described in
(\ref{sextic-potential})  and we can omit the metric term for simplicity). 

Writing down the Schr\"odinger equation of potentials like this, one obtains a
coupled system of differential equations in $\xi$ and $\eta$, respectively,
which are functionally identical. Their difference is that they are defined
on another domain in the
complex plane \cite{KMP6}. If one looks now for bound state solutions, i.e.,
solutions which can be written in terms of polynomials and which are therefore
square-integrable, one finds a quantization condition for the energy
$E$. Because the potential is assumed to be separable in Cartesian coordinates
we already know the energy levels, $E_n$. The second separation constant
$\lambda$ of the system of coupled differential equations in $\xi$ and $\eta$
can then be expressed in terms of $E_n$, i.e. $\lambda_n=f(E_n)$.
The wave-functions of the bound state solutions are determined by three-term
recursion relations, terminating to give polynomials.
However, they cannot be solved to give explicit formulas for the polynomials.

Now we can return to the quasi-exactly solvable potentials. We take one of the
two coupled differential equations and rename the variable $\xi\to x\in\bbbr$,
say. This one-dimensional quasi-exactly solvable potential ''remembers'' its
origin from a two-dimensional superintegrable potential: The subset of
wave-functions which can be explicitly found correspond to the case where one
of the coupling constants corresponds in a simple way with the energy-levels
of the superintegrable potential labeled by $n$, and the emerging
energy-levels of the quasi-exactly solvable potential are determined by the
separation constant $\lambda_n$ of the coupled system of differential
equations. This feature is common to all quasi-exactly solvable potentials,
and even more, one is able to construct quasi-exactly solvable potentials from 
superintegrable potentials in two, three, etc. dimensions.
They are of power-like behavior, or powers of trigonometric, hyperbolic, and
elliptic functions.

However, there does not exist a theory of the corresponding wave-functions,
which are determined by terminating three-term recursion relations for the
bound states and non-terminating three-term recursion relations for the
scattering states. In comparison to the (confluent) hypergeometric
functions little is known about expansion and addition theorems (with the 
exception of Mathieu and spheroidal wave-functions in flat space \cite{MESCH}).
In some few cases, an interbasis expansion is known to switch
from, say, Hermite polynomials to these new wave-functions \cite{KMP6}.

Summarizing, we are not able to treat systems with the structure of 
(\ref{sextic-potential}), and similar with powers of trigonometric and
hyperbolic functions any further. 

\subsubsection{Separation of $V_1$ in the $(u,v)$-System.}
We insert $V_1$ in (\ref{PathIntegralDI}) and obtain
\begin{eqnarray}
&&K^{(V_1)}(u'',u',v'',v';T)
=\pathint{u}\pathint{v}2u
\nonumber\\   &&\qquad\times
\exp\left\{\ih \int_0^T\left[mu(\dot u^2+\dot v^2)
-\frac{1}{2u}\left(\frac{m}{2}\omega^2(4u^2+v^2)+\kappa+
\frac{\lambda^2-\viert}{2mv^2}\right)\right]\dt\right\}
\nonumber\\   &&
=\sqrt{v'v''}\sum_{n=0}^\infty
\Psi_n^{(RHO,\lambda)}(v'')\Psi_n^{(RHO,\lambda)}(v')
K_n^{(V_1)}(u'',u';T)\enspace,
\label{PathIntegralDIV1}
\end{eqnarray}
with the path integral $K_n(T)$ given by
\begin{eqnarray}
&&\!\!\!\!\!\!\!\!
K_n^{(V_1)}(u'',u';T)
\nonumber\\   &&\!\!\!\!\!\!\!\!
=(4u'u'')^{1/4}
\pathint{u}\sqrt{2u}\exp\left\{\ih\int_0^T\left[mu\dot u^2
-\frac{1}{2u}\Big(m\omega^2u^2+\kappa\Big)
-\frac{E_n}{2u}\right]\dt\right\}\enspace,\qquad
\label{PathIntegralDIV1n}
\end{eqnarray}
with $E_n=\hbar\omega(2n+\lambda+1)$ and we have inserted the path integral
solution for the radial harmonic oscillator (RHO) with parameter $\lambda$ 
and the variable $v>0$. If $v$ is more restricted, say $v$ is an angular
variable, additional boundary conditions must be imposed. However,
we continue with the case $v>0$.
The wave-functions for the radial harmonic oscillator 
$V(r)=\frac{m}{2}\omega^2-\frac{\hbar^2}{2m}\frac{\lambda^2-1/4}{r^2}$
have the form
\begin{equation}
\Psi_n^{(RHO,\lambda)}(r)
=\sqrt{\frac{2m}{\hbar}\frac{n!}{\Gamma(n+\lambda+1)}\,r}
  \bigg({m\omega\over\hbar}r\bigg)^{\lambda/2}
  \exp\bigg(-{m\omega\over2\hbar}r^2\bigg)
  L_n^{(\lambda)}\bigg({m\omega\over\hbar}r^2\bigg)
\end{equation}
The $L_n^{(\lambda)}(z)$ are Laguerre polynomials \cite{GRA}.

In the next step we perform a space-time transformation in 
(\ref{PathIntegralDIV1n}) by eliminating the term $2u$ in the metric. 
This gives in the usual way
\begin{equation}
G_{n}^{(V_1)}(u'',u';E)=\int_0^\infty\d s''
\exp\bigg[\ih \bigg(\frac{E^2}{2m\omega^2}-\kappa-E_n\bigg)s''\bigg]
K_n^{(V_1)}(u'',u';s'')\enspace,
\end{equation}
with the transformed path integral given by
\begin{equation}
K_n^{(V_1)}(u'',u';s'')
=\pathints{u}\exp\Bigg\{\ih\ints \Bigg[\frac{m}{2}\dot u^2
-\frac{m}{2}(2\omega)^2\bigg(u-\frac{E}{m\omega}\bigg)^2\Bigg]\d s'\Bigg\}.
\end{equation}
This path integral of a shifted harmonic oscillator with frequency $2\omega$
can be solved. The corresponding Green function has the form
\begin{equation}
G_u^{(V_1)}(E;u'',u';\CE)=\sqrt{\frac{m}{2\pi\hbar^3\omega}}\,
\Gamma\bigg(\half-\frac{\CE}{2\hbar\omega}\bigg)
D_{-\half+\CE/2\hbar\omega}
\left(\sqrt{\frac{4m\omega}{\hbar}}\,\tilde u_>\right)
D_{-\half+\CE/2\hbar\omega}
\left(-\sqrt{\frac{4m\omega}{\hbar}}\,\tilde u_<\right).
\end{equation}
Here, the $D_\nu(z)$ are parabolic cylinder functions \cite{GRA} and $\tilde u=
u-E/2m\omega^2$.
For the evaluation of the $s''$-integration we use the involution formula
\begin{equation}
G(u'',u',v'',v';E)
=\frac{\hbar}{2\pi\i}\int\d\CE G_v(E;v'',v';\CE) G_u(E;u'',u';-\CE)\enspace.
\end{equation}
to obtain
\begin{eqnarray}
K^{(V_1)}(u'',u',v'',v';T)
&=&\int_{-\infty}^\infty\frac{\d E}{2\pi\hbar}\,\e^{-\i ET/\hbar}
\sqrt{v'v''}\sum_{n=0}^\infty\Psi^{(RHO,\lambda)}(v'')\Psi^{(RHO,\lambda)}(v')
\nonumber\\  &&\qquad\times
G_u^{(V_1)}
\bigg[E;u'',u';\bigg(\frac{E^2}{2m\omega^2}-\kappa-E_n\bigg)\bigg]\enspace.
\end{eqnarray}

\subsubsection*{Solution without Boundary Condition}
Let us first solve the potential problem $V_1$ on $\DI$ without any boundary
condition on the variables. In this case the path integral in the variable $u$
is just a path integral for a shifted harmonic oscillator with wave-functions
given by $\Psi_l^{(HO)}(\tilde u)$  with $\tilde u=u-E/m\omega^2$. 
The wave-functions for the harmonic oscillator (HO) are given by the
well-known form in terms of Hermite-polynomials
\begin{equation}
\Psi_n^{(HO)}(x)=
  \bigg({m\omega\over\pi\hbar}{1\over2^nn!}^{1/2}\bigg)^{1/4}
   H_n\bigg(\sqrt{m\omega\over\hbar}\,x\bigg)
   \exp\bigg(-{m\omega\over2\hbar}x^2\bigg)\enspace.
\end{equation}
Evaluating the Green function $G_u^{(V_1)}$ we obtain the solution:
\begin{eqnarray}
K^{(V_1)}_{\rm discr.}(u'',u',v'',v';T)
&=&
\sum_{n=0}^\infty\sum_{l=0}^\infty
\sqrt{\frac{m\omega^2}{2E_{ln}}\,v'v''}\,\e^{-\i E_{ln}T/\hbar}
\nonumber\\  &&\qquad\times
\Psi_n^{(RHO,\lambda)}(v'')\Psi_n^{(RHO,\lambda)}(v')
\Psi_l^{(HO)}(\tilde u'')\Psi_l^{(HO)}(\tilde u')
\enspace,
\\  
E_{ln}&=&\pm\sqrt{m\hbar\omega^3(2l+2n+2+\lambda)+2m\omega^2\kappa}\enspace.
\label{Eln-V1-DI}
\end{eqnarray}
The spectrum is degenerate in $n$ and $l$, as it is known for
superintegrable potentials.
However, this ''solution'' is seriously flawed. If we calculate the norm of 
the wave-functions, we see immediately that the norm is proportional to the
energy $E_n$, which in the negative-sign case is negative, 
and it follows that the Hilbert space is not properly defined.
In the positive-sign case the norm would be positive, however, the
corresponding configuration space cannot be extended to $u\to-\infty$,
and this does not make sense either.

\subsubsection*{Solution with Boundary Condition}
Due to the coordinate singularity for $u=0$ we must impose some boundary
condition. The simplest way to incorporate such a boundary condition is to
require that the wave-functions vanish at $u=0$, or generally the motion in
the variable $u$ takes place only in the half-space $u>a$. By exploiting the
Dirichlet  boundary-conditions \cite{GROr} at $u=a$ we therefore get
\begin{eqnarray}
&&G_{(x=a)}^{(V_1)}(u'',u',v'',v';E)=
\sqrt{v'v''}\sum_{n=0}^\infty
\Psi_n^{(RHO,\lambda)}(v'')\Psi_n^{(RHO,\lambda)}(v'')\qquad
\nonumber\\   &&\qquad\qquad\qquad\qquad\times
\left\{G_n^{(V_1)}(u'',u';E)-\frac{G_n^{(V_1)}(u'',a;E)G_n^{(V_1)}(a,u';E)}
{G_n^{(V_1)}(a,a;E)}\right\}\enspace.
\label{GhalfspaceV1}
\end{eqnarray}
This Green function cannot be evaluated further. However, we can determine
bound states by the poles of (\ref{GhalfspaceV1}) and obtain the 
quantization condition
\begin{eqnarray}
&&D_{\nu_{l,n}}\left[2\sqrt{\frac{m\omega}{\hbar}}\,
\bigg(a-\frac{E_{l,n}}{m\omega^2}\bigg)\,\right]=0\enspace,
   \\   &&
\nu_{l,n}=-\half+\frac{1}{2\omega\hbar}
\bigg(\frac{E_{l,n}^2}{m\omega^2}-\kappa-\hbar\omega(2n+\lambda+1)
\bigg)\enspace.
\label{Dnu-V1-DI}
\end{eqnarray}
According to \cite{KalninsKWinter} the asymptotic behavior of the 
energy-eigenvalues is in accordance with (\ref{Eln-V1-DI}) for high-level
states. The wave-functions can be obtained by taking the residuum of the 
curly-bracket expression in (\ref{GhalfspaceV1}).

Our last quantization condition, however, rises a problem.
It is not obvious for us how to determine the degeneracy of the 
energy-values which is usually typically for superintegrable systems. 
The solution (\ref{Eln-V1-DI}) has this degeneracy but the boundary 
conditions are not fulfilled and the Hilbert space is not properly 
defined either.
For the solution (\ref{Dnu-V1-DI}) it is just the other way
round. In the original paper \cite{KalninsKWinter} this issue was not
addressed any further. 

We can see from the quantization condition (\ref{Dnu-V1-DI}) that for each
value of the number $n$ a set of energy levels $E_{l,n}$ follows, i.e.
a set $E_{l,0},E_{l,1},\dots$. There is no possibility to find that a level
from the set $n=0$ is equal to one level of the set $n=1$, for example
$E_{l_a,0}=E_{l_b,1}$ for some numbers $l_a,l_b$.
Therefore we find that the degeneracy of the energy levels is lost.
The usual lore in the study of superintegrable systems is that the 
statements that a potential is superintegrable and that the spectrum 
of such a potential is degenerate are equivalent.
Indeed, from the Sturm-Liouville theory for differential equations,
i.e. in our case the quantum Hamiltonian, it follows that degeneracy
implies superintegrability, i.e. additional constants of motion.
However, this statement is not valid the other way round, and the present
examples of potentials on Darboux space $\DI$ serve as counter examples
for such an attempt. 

If we look at (\ref{GhalfspaceV1}) we see that the ''lost'' degeneracy
is due to the boundary condition for the Green function and the 
wave-functions, respectively, for some $u>a>0$.
For $u=0$ the curvature of the space becomes infinite and a wave-function
at the coordinate origin does not make sense.
Depending whether the Darboux space  $\DI$ is embedded in three-dimensional
space with definite or indefinite metric further determines the parameter $a$,
c.f. \cite{KalninsKWinter}. For a positive-definite metric, $v$ is an angle
with $v \in[0,2 \pi)$, the two dimensional surface making up  $\DI$ has a 
definite boundary and it follows $a=1/2$.
For a negative-definite metric the boundary turns out to be constraint by
$a=0$. In fact, it is not possible to extend the surface beyond $u<0$ and all values from 0 to $\infty$ are definitely excluded.
We will see that the same property holds for the potential $V_2$.

\subsection{The Superintegrable Potential $V_2$ on $\DI$.}
Next, we consider the potential $V_2$ on on $\DI$.  First, we state
the potential in the separating coordinate systems. We have
\begin{eqnarray}
V_2(u,v)&=&\frac{1}{2u}\bigg[\frac{m}{2}\omega^2(u^2+v^2)+\kappa_1+
\kappa_2v\bigg]\enspace,
   \\   &=&\frac{1}{2u}
\bigg[\frac{m}{2}\omega^2(r^2+q^2)+\kappa_1+
\kappa_2(q\cos\vtheta-r\sin\vtheta)\bigg]\enspace.
\label{V2rq-system}
\end{eqnarray}

\subsubsection{Separation of $V_2$ in the $(u,v)$-System.}
We proceed in a similar way as before and obtain
\begin{eqnarray}
&&K^{(V_2)}(u'',u',v'',v';T)
=\pathint{u}\pathint{v}2u
\nonumber\\   &&\qquad\times
\exp\left\{\ih \int_0^T\left[mu(\dot u^2+\dot v^2)
-\frac{1}{2u}\left(\frac{m}{2}\omega^2(u^2+v^2)+\kappa_1+
+\kappa_2v\right)\right]\dt\right\}
\nonumber\\   &&
=\sum_{n=0}^\infty
\Psi_n^{(HO)}(\tilde v'')\Psi_n^{(HO)}(\tilde v')
K_n^{(V_2)}(u'',u';T)\enspace,
\label{PathIntegralDIV2}
\end{eqnarray}
where $\Psi_n^{(HO)}$ are the wave-functions of a shifted
harmonic oscillator with $\tilde v=v+\kappa_2/m\omega$. 
Note that we have to require $v\in\bbbr$, otherwise for $v$ cyclic
complicated boundary conditions have to imposed on the solution
in $v$. The remaining path integral in the variable $u$ has the form
\begin{eqnarray}
&&K_n^{(V_2)}(u'',u';T)
=(4u'u'')^{1/4}\pathint{u}\sqrt{2u}
\nonumber\\   &&\qquad\times
\exp\left\{\ih\int_0^T\left[mu\dot u^2
-\frac{1}{2u}\left(\frac{m}{2}\omega^2u^2+\kappa_1+\hbar(n+\bhalf)
-\frac{\kappa_2^2}{2m\omega^2}\right)\right]\dt\right\}.
\end{eqnarray}
This gives in the usual way
\begin{equation}
G_n^{(V_2)}(u'',u';E)=\int_0^\infty\d s''
\exp\bigg[-\ih s''\bigg(
\kappa_1+\hbar\omega(n+\bhalf)-\frac{\kappa_2^2}{2m\omega^2}\bigg)\bigg]
K_n^{(V_2)}(u'',u';s'')\enspace,
\end{equation}
with the transformed path integral given by ($\tilde u=u-2E/m\omega^2$)
\begin{eqnarray}
K_n^{(V_2)}(u'',u';s'')
&=&\pathints{u}\exp\Bigg\{\ih\ints \Bigg[
\frac{m}{2}(\dot u^2-\omega^2 u^2)+2Eu\Bigg]\d s'\Bigg\}
\nonumber\\   &=&
\e^{2\i s''E/m\omega^2\hbar}
\pathints{u}\exp\Bigg[\frac{\i m}{2\hbar}\ints \frac{m}{2}
(\dot{\tilde u}^2-\omega^2\tilde u^2)\d s'\Bigg].\qquad
\end{eqnarray}

\subsubsection*{Solution without Boundary Condition}
This is again a path integral for a shifted harmonic oscillator, and first
we ignore the boundary condition for the wave-functions in the variable $u$
for $u=0$, say, we obtain the solution:
\begin{eqnarray}
K^{(V_2)}_{\rm discr.}(u'',u',v'',v';T)
&=&
\sum_{n=0}^\infty\sum_{l=0}^\infty
\sqrt{\frac{m\omega^2}{4E_{ln}}v'v''}\,\e^{-\i E_{ln}T/\hbar}
\nonumber\\  &&\quad\times
\Psi_n^{(HO)}(\tilde v'')\Psi_n^{(HO)}(\tilde v')
\Psi_n^{(HO)}(\tilde u'')\Psi_n^{(HO)}(\tilde u')
\enspace,\qquad
\\  
E_{ln}&=&\pm\sqrt{\frac{m\hbar\omega^2}{2}
\bigg(l+n+1+\kappa_1-\frac{k_2^2}{2m\omega^2}\bigg)}\enspace.
\label{Eln-V2-DI}
\end{eqnarray}
This spectrum exhibits degeneracy, however the norm is again proportional to 
the energy, which is negative, and therefore the Hilbert space is not 
properly defined..

\subsubsection*{Solution with Boundary Condition}
If we now take into account the boundary condition for some $u=a$ such that
the wave-function vanish for $u=a$, we obtain in a similar manner as in the
previous subsection:
\begin{eqnarray}
&&G_{(x=a)}^{(V_2)}(u'',u',v'',v';E)=
\sqrt{v'v''}\sum_{n=0}^\infty
\Psi_n^{(HO)}(\tilde v'')\Psi_n^{(HO)}(\tilde v')
\nonumber\\   &&\qquad\qquad\qquad\qquad\times
\left\{G_n^{(V_2)}(u'',u';E)-\frac{G_n^{(V_2)}(u'',a;E)G_n^{(V_2)}(a,u';E)}
{G_n^{(V_2)}(a,a;E)}\right\}\enspace,
\end{eqnarray}
with the Green function $G_n^{(V_2)}(E)$ given by
\begin{eqnarray}
G_u^{(V_2)}(E;u'',u';\CE)&=&\sqrt{\frac{m}{2\pi\hbar^3\omega}}\,
\Gamma\bigg(\half-\frac{\CE}{2\hbar\omega}\bigg)
\nonumber\\   &&\qquad\times
D_{-\half+\CE/2\hbar\omega}
\left(\sqrt{\frac{4m\omega}{\hbar}}\,\tilde u_>\right)
D_{-\half+\CE/2\hbar\omega}
\left(-\sqrt{\frac{4m\omega}{\hbar}}\,\tilde u_<\right),\qquad\qquad
\\
\CE&=&\frac{2E^2+\kappa_2^2/2}{m\omega^2}-\kappa_1-\hbar\omega(n+\bhalf)
\enspace.
\label{GhalfspaceV2}
\end{eqnarray}
Bound states can be determined by the quantization condition
\begin{eqnarray}
&&D_{\nu_{l,n}}\left[\sqrt{\frac{2m\omega}{\hbar}}\,
\bigg(a-\frac{2E_{l,n}}{m\omega^2}\bigg)\,\right]=0\enspace,
   \\   &&
\nu_{l,n}=-\half+\frac{1}{\omega\hbar}
\bigg(\frac{2E^2_{ln}+\kappa_2^2/2}{m\omega^2}
-\kappa_1-\hbar\omega(n+\bhalf)\bigg)\enspace.
\label{LevelsV2}
\end{eqnarray}
Again, degeneracy in the quantum numbers $n$ and $l$ is lost. 
According to
\cite{KalninsKWinter} the asymptotic behavior of the  energy-eigenvalues
(\ref{LevelsV2}) is in accordance with (\ref{Eln-V2-DI}).
 The wave-functions can be obtained by taking the residuum of the 
curly-bracket expression in (\ref{GhalfspaceV2}).

\subsubsection{Separation of $V_2$ in the $(r,q)$-System.}
In order to set up the path integral formulation we follow our
canonical procedure. The Lagrangian and Hamiltonian are given by,
respectively: 
\begin{eqnarray}
\CL(r,\dot r, q,\dot q)
=m(r\cos\vtheta+q\sin\vtheta)(\dot r^2+\dot q^2)-V(r,q)\enspace,
\\
\CH(r,p_r,q,p_q)
=\frac{1}{4m(r\cos\vtheta+q\sin\vtheta)}(p_r^2+p_q^2)+V(r,q)\enspace.
\end{eqnarray}
The canonical momenta are
\begin{eqnarray}
p_r&=&\hi\bigg(\frac{\partial}{\partial r}
+\frac{\cos\vtheta}{2(r\cos\vtheta+q\sin\vtheta)}\bigg)\enspace,
\\
p_q&=&\hi\bigg(\frac{\partial}{\partial q}
+\frac{\sin\vtheta}{2(r\cos\vtheta+q\sin\vtheta)}\bigg)\enspace.
\end{eqnarray}
The quantum Hamiltonian has the form
\begin{eqnarray}
H&=&-\frac{\hbar^2}{2m}\frac{1}{2(r\cos\vtheta+q\sin\vtheta)}
\bigg(\frac{\partial^2}{\partial r^2}
 +\frac{\partial^2}{\partial q^2}\bigg)+V(r,q)
\\
&=&\frac{1}{2m}\frac{1}{\sqrt{2(r\cos\vtheta+q\sin\vtheta)}}(p_r^2+p_q^2)
               \frac{1}{\sqrt{2(r\cos\vtheta+q\sin\vtheta)}}+V(r,q)\enspace.
\end{eqnarray}
Using the representation (\ref{V2rq-system}) we write down the path
integral for $V_2$ in the rotated $(r,q)$-coordinate system, and
obtain
\begin{eqnarray}
&&K(r'',r',q'',q';T)=\pathint{r}\pathint{q}
2(r\cos\vtheta+q\sin\vtheta)
\nonumber\\  &&\qquad\times
\exp\Bigg\{\ih\int_0^T\Bigg[m(r\cos\vtheta+q\sin\vtheta)
(\dot r^2+\dot q^2)
\nonumber\\  &&\qquad\qquad\qquad\qquad\qquad\qquad
-\frac{1}{2u}
\bigg(\frac{m}{2}\omega^2(r^2+q^2)+\kappa_1+
\kappa_2(q\cos\vtheta-r\sin\vtheta)\bigg)\Bigg]\dt\Bigg\}.\qquad
\end{eqnarray}
Performing a space-time transformation in the usual way gives
\begin{equation}
G(r'',r',q'',q';E)=
\int_0^\infty \d s''\e^{-\i s''\kappa_1/\hbar}
   K(r'',r',q'',q';s'')\enspace,
\end{equation}
with the transformed path integral $K(s'')$ given by
\begin{eqnarray}
&&K(r'',r',q'',q';s'')
=\pathints{r}\pathints{q}
\exp\Bigg\{\ih\ints \Bigg[
\frac{m}{2}\Big(\dot r^2+\dot q^2-\omega^2(r^2+q^2)\Big)
\nonumber\\  &&\qquad\qquad\qquad\qquad\qquad\qquad
+2E(r\cos\vtheta+q\sin\vtheta)
-\kappa_2(-r\sin\vtheta+q\cos\vtheta)\Bigg]\d s\Bigg\}
\nonumber\\  &&
=\exp\Bigg[\ih\Bigg(\frac{4E^2+\kappa_2^2}{2m\omega^2}-\kappa_1\Bigg)s''\Bigg]
\exp\left\{\ih\ints \left[
\frac{m}{2}\Big(\dot r^2+\dot q^2)-\frac{m}{2}\omega^2(\tilde
r^2+\tilde q^2)\right]\d s\right\}
\nonumber\\  &&
=\exp\Bigg[\ih\Bigg(\frac{4E^2+\kappa_2^2}{2m\omega^2}
-\hbar\omega(n+\bhalf)-\kappa_1\Bigg)s''\Bigg]
\sum_{n=0}^\infty
\Psi_n^{(HO)}(\tilde q'')\Psi_n^{(HO)}(\tilde q')
\nonumber\\  &&\qquad\qquad\qquad\qquad\times
\pathints{r}\exp\left[\frac{\i m}{2\hbar}\ints 
(\dot r^2-\omega^2\tilde r^2)\d s\right]
\end{eqnarray}
($\tilde r=r-(2E\cos\vtheta+\kappa_2\sin\vtheta)/m\omega^2$,
$\tilde q=q-(2E\sin\vtheta-\kappa_2\cos\vtheta)/m\omega^2$.)
Here, we have inserted the path integral solution for the shifted
harmonic oscillator in the variable $q$. 


\subsubsection*{Solution with Boundary Condition}
For the path integral for the shifted harmonic oscillator in the
variable $r$ we now take care that the variable $u$ is defined only in
the half-space $u\geq a$. Setting for instance in the definition of
the $(r,q)$-system $\vtheta=0$ yields $r=u$ and $q=v$. For
$\vtheta=\pi/2$ the roles of $r$ and $q$ are reversed. In the view of
the previous paragraph of $V_2$ in the $(u,v)$-system we impose of 
the Green function in $r$ the boundary condition $r\geq a$ and obtain
in this limiting case for the bound states the quantization condition
\begin{eqnarray}
&&D_{\nu_{l,n}}\left[\sqrt{\frac{2m\omega}{\hbar}}\,
\bigg(a-\frac{2E_{l,n}}{m\omega^2}\bigg)\,\right]=0\enspace,
   \\   &&
\nu_n=-\half+\frac{1}{\omega\hbar}
\bigg(\frac{2E^2_{ln}+\kappa_2^2/2}{m\omega^2}
-\kappa_1-\hbar\omega(n+\bhalf)\bigg)\enspace.
\label{LevelsV2b}
\end{eqnarray}
This is the result of (\ref{LevelsV2}). The quantization conditions of 
(\ref{LevelsV2}) and (\ref{LevelsV2b}) are identical as it should be.

\subsection{The Superintegrable Potential $V_3$ on $\DI$.}
Next, we consider the potential $V_3$ on on $\DI$.  First, we state
the potential in the separating coordinate systems. We have
\begin{eqnarray}
V_3(u,v)&=&\frac{1}{2u}\frac{\hbar^2v_0^2}{2m},
   \\   &=&\frac{1}{\xi^2-\eta^2+2c}\frac{\hbar^2v_0^2}{2m},
   \\   &=&\frac{1}{2(r\cos\vtheta+q\sin\vtheta)}\frac{\hbar^2v_0^2}{2m}.
\end{eqnarray}
This potential can be considered as a special case either of $V_1$ or
$V_2$, respectively. However, it has an additional conserved quantum
number, i.e. $K=p_v$. Therefore we will sketch only the solution in
the $(u,v)$-system. Proceeding in the usual way, we obtain for the
path integral (assuming $v$ cyclic):
\begin{eqnarray}
&&\!\!\!\!\!\!\!\!\!
K(u'',u',v'',v';T)
=\pathint{u}\pathint{v}2u\exp\left\{\ih\int_0^T
\left[mu(\dot u^2+\dot v^2)-\frac{1}{2u}\frac{\hbar^2v_0^2}{2m}\right]
\dt\right\}
\nonumber\\   &&\!\!\!\!\!\!\!\!\!
         \\   &&\!\!\!\!\!\!\!\!\!
=(4u'u'')^{1/4}\sum_{l=0}^\infty\frac{\e^{\i l(v''-v')}}{2\pi}
\pathint{u}\sqrt{2u}
\exp\left\{\ih\int_0^T
\left[mu\dot u^2-\frac{1}{2u}\frac{\hbar^2}{2m}(l^2+v_0^2)\right]
\dt\right\}.
\end{eqnarray}
We observe that the only effect is change in the quantum number $l$ in
comparison to the $v_0=0$ case. Using the solution of \cite{GROas} we
get for the corresponding Green function
\begin{eqnarray}
&&G(u'',u',v'',v';E)=
\sum_{l=-\infty}^\infty\frac{\e^{\i l(v''-v')}}{2\pi}
\frac{4m}{3\hbar}
\bigg[\bigg(u'-\frac{\tilde l^2\hbar^2}{4mE}\bigg)
\bigg(u''-\frac{\tilde l^2\hbar^2}{4mE}\bigg)\bigg]^{1/2}
\nonumber\\  &&\qquad\times
\left[\tilde I_{1/3}\bigg(u_<-\frac{\tilde l^2\hbar^2}{4mE}\bigg)
      \tilde K_{1/3}\bigg(u_>-\frac{\tilde l^2\hbar^2}{4mE}\bigg)
\vphantom{\dfrac{\tilde I_{1/3}\bigg(a-\frac{\tilde l^2\hbar^2}{4mE}\bigg)}
      {\tilde K_{1/3}\bigg(a-\frac{\tilde l^2\hbar^2}{4mE}\bigg) }}\right. 
\nonumber\\  &&\qquad\qquad\qquad\qquad\left.
-\dfrac{\tilde I_{1/3}\bigg(a-\frac{\tilde l^2\hbar^2}{4mE}\bigg)}
      {\tilde K_{1/3}\bigg(a-\frac{\tilde l^2\hbar^2}{4mE}\bigg) }
\tilde K_{1/3}\bigg(u'-\frac{\tilde l^2\hbar^2}{4mE}\bigg)
\tilde K_{1/3}\bigg(u''-\frac{\tilde l^2\hbar^2}{4mE}\bigg)\right]
\enspace.\qquad\qquad
\label{G-DarbouxI}
\end{eqnarray}
$\tilde I_\nu(z)$ denotes
$$
\tilde I_\nu(z)=I_\nu\bigg(\frac{4\sqrt{-mE}}{3\hbar}z^{3/2}\bigg)\enspace,
$$
with $\tilde K_\nu(z)$ similarly, and $\tilde l^2=l^2+v_0^2$.
Due to the relation of the Airy-function \cite{ABS,GRA}
$K_{\pm 1/3}(\zeta)=\pi\sqrt{3/z}\, \Ai(z)$, $z=(3\zeta/2)^{2/3}$, and 
the observation that for $E<0$ the argument of $\Ai(z)$ is always greater
than zero, and there are no bound states. For $E>0$ there is no real 
bound state solution, either. This concludes the discussion.


\setcounter{equation}{0}
\section{Superintegrable Potentials on Darboux Space $\DII$}
\message{Superintegrable Potentials on Darboux Space D_II}
In this section we consider superintegrable potentials in 
the Darboux Space $\DII$ (\ref{DarbouxII}).
The following four coordinate systems separate the Schr\"odinger equation
for the free motion:
\begin{eqnarray}
\hbox{($(u,v)$-System:)}&& x=\half(v+\i u),\quad y=\half(v-\i u)\,,
\\
\hbox{(Polar:)}&& u=\vrho\cos\vtheta,\quad v=\vrho\sin\vtheta\enspace,\quad
\qquad\qquad\quad\,\,\,\,\,(\vrho>0,
\vtheta\in(-\hbox{$\frac{\pi}{2}$},\hbox{$\frac{\pi}{2}$}))\,,
\\
\hbox{(Parabolic:)}&& u=\xi\eta,\quad v=\half(\xi^2-\eta^2)\enspace,\quad 
\qquad\qquad\quad\,\,\,\,(\xi>0,\eta>0)\,,
\\
\hbox{(Elliptic:)}&& u=d\cosh\omega\cos\vphi,\quad 
                     v=d\sinh\omega\sin\vphi\enspace,\quad
(\omega>0,\vphi\in(-\hbox{$\frac{\pi}{2}$},\hbox{$\frac{\pi}{2}$}))\,.
\qquad
\end{eqnarray}
$2d$ is the interfocal distance in the elliptic system. For convenience 
we also display in the following the special case of the parameters
$a=-1$ and $b=1$ \cite{KalninsKMWinter}. The infinitesimal distance
is given in these four cases (note that the metric gives us the additional
requirement $u>0$):
\begin{eqnarray}
\d s^2 &=&\frac{bu^2-a}{u^2}(\d u^2+\d v^2)
\enspace,
\label{metric-uv}
\\
\hbox{(Polar:)}
&=&\frac{b\vrho^2\cos^2\vtheta-a}{\vrho^2\cos^2\vtheta}
(\d\vrho^2+\vrho^2\d\vtheta^2)
\enspace,
\\
\hbox{(Parabolic:)}
&=&\frac{b\xi^2\eta^2-a}{\xi^2\eta^2}(\xi^2+\eta^2)(\d\xi^2+\d\eta^2)
\nonumber\\
&=&\bigg[\bigg(b\xi^2-\frac{a}{\xi^2}\bigg)
+\bigg(b\eta^2-\frac{a}{\eta^2}\bigg)\bigg](\d\xi^2+\d\eta^2)\enspace, 
\\
\hbox{(Elliptic:)}
&=&\frac{bd^2\cosh^2\omega\cos^2\vphi-a}{\cosh^2\omega\cos^2\vphi}
        (\cosh^2\omega-\cos^2\vphi)(\d\omega^2+\d\vphi^2)\enspace,
\nonumber\\
&=&\bigg[\bigg(bd^2\cosh^2\omega+\frac{a}{\cosh^2\omega}\bigg)\!-\!
\bigg(bd^2\cos^2\vphi+\frac{a}{\cos^2\vphi}\bigg)\bigg]
(\d\omega^2+\d\vphi^2)
\,.\,\qquad
\end{eqnarray}
We can see that the case $a=-1$, $b=0$ leads to the case
of the Poincar\'e upper half-plane $u>0$ endowed with the
metric (\ref{metric-uv}) \cite{GROad}, i.e. the
two-dimensional hyperboloid $\Lambda^{(2)}$ in horicyclic coordinates. 
The parabolic case corresponds to the semi-circular-parabolic system
and the elliptic case to the elliptic-parabolic system on the two-dimensional
hyperboloid. On the other hand, the case $a=0$, $b=1$ just gives the usual
two-dimensional Euclidean plane with its four coordinate system which allow
separation of variables of the Laplace-Beltrami equation, i.e., the
Cartesian, polar, parabolic, and elliptic system. Hence, the Darboux space II
contains as special cases a space of constant zero curvature (Euclidean
plane) and a space of constant negative curvature (the hyperbolic plane).
This includes the emerging of
coordinate systems in flat space from curved spaces. 

\begin{table}[t!]
\caption{\label{cosytabDII} 
Constants of Motion and Limiting Cases of Coordinate Systems on $\DII$}
\hfuzz=20pt
\begin{eqnarray}\begin{array}{l}\vbox{\small\offinterlineskip
\halign{&\vrule#&$\strut\ \hfil\hbox{#}\hfill\ $\cr
\noalign{\hrule}
height2pt&\omit&&\omit&&\omit&&\omit&&\omit&\cr
&Metric:&&Constant                     &&$\DII$  
        &&$\Lambda^{(2)} ($a=-1,b=0$)$ && $E_2$ ($a=0,b=1$)&\cr
&       && of Motion                   &&  && && &\cr
height2pt&\omit&&\omit&&\omit&&\omit&&\omit&\cr
\noalign{\hrule}\noalign{\hrule}
height2pt&\omit&&\omit&&\omit&&\omit&&\omit&\cr
&$\dfrac{bu^2-a}{u^2}(\d u^2+\d v^2)$  &&$K^2$
        &&$(u,v)$-System     &&Horicyclic    &&Cartesian     &\cr
height2pt&\omit&&\omit&&\omit&&\omit&&\omit&\cr
\noalign{\hrule}
height2pt&\omit&&\omit&&\omit&&\omit&&\omit&\cr
&$\dfrac{b\vrho^2\cos^2\vtheta-a}{\vrho^2\cos^2\vtheta}(\d\vrho^2+\d\vtheta^2)$
                                       &&$X_2$
         &&Polar       &&Equidistant   &&Polar         &\cr
height2pt&\omit&&\omit&&\omit&&\omit&&\omit&\cr
\noalign{\hrule}
height2pt&\omit&&\omit&&\omit&&\omit&&\omit&\cr
&$\dfrac{b\xi^2\eta^2-a}{\xi^2\eta^2}(\xi^2+\eta^2)(\d\xi^2+\d\eta^2)$
                                       &&$X_1$
         &&Parabolic   &&Semi-circular &&Parabolic     &\cr
&                                      && &&  &&parabolic && &\cr
height2pt&\omit&&\omit&&\omit&&\omit&&\omit&\cr
\noalign{\hrule}
height2pt&\omit&&\omit&&\omit&&\omit&&\omit&\cr
&$\dfrac{bd^2\cosh^2\omega\cos^2\vphi-a}{\cosh^2\omega\cos^2\vphi}$
                                             && && && && &\cr
&$\times(\cosh^2\omega-\cos^2\vphi)(\d\omega^2+\d^2\vphi^2)$
        &&$X_2+d^2 K^2$  &&Elliptic   &&Elliptic-parabolic &&Elliptic &\cr
height2pt&\omit&&\omit&&\omit&&\omit&&\omit&\cr
\noalign{\hrule}}}\end{array}\nonumber\end{eqnarray}
\end{table}
\hfuzz=5pt
\noindent
We find for the Gaussian curvature in the $(u,v)$-system
\begin{equation}
G=\frac{a(a-3bu^2)}{(a-2bu^2)^3}\enspace.
\end{equation}
For $b=0$ we have $G=1/a$ which is indeed a space of constant curvature, 
and the quantity $a$ measures the curvature. In particular, for the 
unit-two-dimensional hyperboloid we have $G=1/a$, with $a=-1$ as the
special case of $\Lambda^{(2)}$. In the following we will assume that
$a<0$ in order to assure the positive definiteness of the metric (1.2).

The following constants of motion are introduced on $\DII$ (without
potential):  
\begin{eqnarray}
K&=&p_v
       \\
X_1&=&\frac{2v(p_v^2-u^2p_u^2)}{bu^2-a}+2up_up_v\enspace,
       \\
X_2&=&\frac{(v^2-u^4)p_v^2+u^2(1-v^2)p_u^2}{bu^2-a}+2uvp_up_v\enspace.
\end{eqnarray}
They satisfy the Poisson algebra relations
\begin{equation}
\{K,X_1\}=2(K^2-\tilde\CH_0)\enspace,\qquad
\{K,X_2\}=X_1\enspace,\qquad
\{X_1,X_2\}=4KX_2\enspace,\qquad
\end{equation}
and the relation 
\begin{equation}
X_1^2-4K^2X_2+4\tilde\CH_0X_2-4\tilde\CH_0^2=0\enspace.
\end{equation}
The quantum analogues have the form (again with $\i$, $\hbar$, $2m$)
\begin{eqnarray}
K&=&\partial_v
       \\
X_1&=&\frac{2v}{bu^2-a}(\partial_v^2-u^2\partial_u^2)
+2u\partial_u\partial_v\enspace,
       \\
X_2&=&\frac{1}{bu^2-a}\bigg[
(v^2-u^4)\partial_v^2+u^2(1-v^2)\partial_u^2\bigg]
+2uv\partial_u\partial_v+u\partial_u+v\partial_v-\bviert\enspace,\qquad
\end{eqnarray}
and satisfy the operator relation ($\widehat H_0$ the Hamiltonian operator,
$\{,\}$ the anti-commutator)
\begin{equation}
\widehat X_1^2-2\{\widehat K^2,\widehat X_2\}+4\widehat H_0\widehat X_2
-4\widehat H_0^2+4\widehat K^2=0\enspace,
\end{equation}
and the commutation relations
\begin{equation}
[\widehat K,\widehat X_1]=2(\widehat K^2-\widehat H_0)\enspace,\qquad
[\widehat K,\widehat X_2]=\widehat X_1\enspace,\qquad
[\widehat X_1,\widehat X_2]=2\{\widehat K,\widehat X_2\}\enspace.\qquad
\end{equation}

\noindent
We consider the following potentials on $\DII$:
\begin{eqnarray}
V_1(u,v)&=&\frac{bu^2-a}{u^2}
\left[\frac{m}{2}\omega^2(u^2+4v^2)+k_1v+\frac{\hbar^2}{2m}
\frac{k_2^2-\viert}{u^2}\right]\enspace,
\\
V_2(u,v)&=&\frac{bu^2-a}{u^2}\left[\frac{m}{2}\omega^2(u^2+v^2)
+\frac{\hbar^2}{2m}\bigg(\frac{k_1^2-\viert}{u^2}
+\frac{k_2^2-\viert}{v^2}\bigg)\right]\enspace,
\\
V_3(u,v)&=&\frac{bu^2-a}{u^2}\frac{2m}{\sqrt{u^2+v^2}}
\left[-\alpha+\frac{\hbar^2}{2m}\left(\frac{k_1^2-\viert}{\sqrt{u^2+v^2}+v}
+\frac{k_2^2-\viert}{\sqrt{u^2+v^2}-v}\right)\right]\enspace,
\\
V_4(u,v)&=&\frac{bu^2-a}{u^2}\frac{\hbar^2}{2m}v_0^2\enspace.
\end{eqnarray}
In the Table \ref{PotentialsDII} we have listed the properties of these
potentials (the coordinate systems where an explicit path integral evaluation
is possible  are $\underline{\hbox{underlined}}$).

\hfuzz=30pt
\begin{table}[t!]
\caption{Separation of variables for the
  superintegrable potentials on $\DII$}
\label{PotentialsDII}
\hfuzz=30pt
\begin{eqnarray}\begin{array}{l}\vbox{\small\offinterlineskip
\halign{&\vrule#&$\strut\ \hfil\hbox{#}\hfill\ $\cr
\noalign{\hrule}
height2pt&\omit&&\omit&&\omit&&\omit&&\omit&\cr
&Potential&&Constants of Motion &&Separating coordinate &\cr
&         &&                    &&system &\cr
height2pt&\omit&&\omit&&\omit&&\omit&&\omit&\cr
\noalign{\hrule}\noalign{\hrule}
height2pt&\omit&&\omit&&\omit&&\omit&&\omit&\cr
&$V_1$    &&$R_1=X_1+m\omega^2v\Big(u^2+\frac{u^2+4v^2}{bu^2-a}\Big)
                 +\frac{k_1}{2}\Big(u^2+\frac{4v^2}{bu^2-a}\Big)
                 -\hbar^2\frac{k_2^2-\viert}{m}\frac{v}{bu^2-a}$
            &&$\underline{\hbox{$(u,v)$-System}}$ &\cr
&         &&$R_2=K^2+2m\omega^2v^2+k_1v$
            &&Parabolic                           &\cr
height2pt&\omit&&\omit&&\omit&&\omit&&\omit&\cr
\noalign{\hrule}
height2pt&\omit&&\omit&&\omit&&\omit&&\omit&\cr
&$V_2$    &&$R_1=X_2+\frac{u^2+v^2}{bu^2-a}\Big[\frac{m}{2}\omega^2(u^2+v^2)
                    -\frac{\hbar^2}{2m}\Big(k_1^2-\viert
                     -(k_2^2-\viert)\frac{u^2}{v^2}\Big)\Big]$
              &&$\underline{\hbox{$(u,v)$-System}}$ &\cr
&         &&$R_2=K^2+\frac{m}{2}\omega^2v^2
                 +\frac{\hbar^2}{2m}\frac{k_2^2-\viert}{v^2}$
              &&$\underline{\hbox{Polar}}$          &\cr
&         &&  &&Elliptic                            &\cr
height2pt&\omit&&\omit&&\omit&&\omit&&\omit&\cr
\noalign{\hrule}
height2pt&\omit&&\omit&&\omit&&\omit&&\omit&\cr
&$V_3$    &&$R_1=X_1+\frac{-\alpha\xi^2(\eta^4+1)
                    +\frac{\hbar^2}{2m}(k_1^2-\viert)(\eta^4+1)
                    -\frac{\hbar^2}{2m}(k_2^2-\viert)(\xi^4+1)}
                           {(b\xi^2\eta^2-a)(\xi^2+\eta^2)}$
              &&$\underline{\hbox{Polar}}$          &\cr
&         &&$R_2=X_2-\frac{\alpha(\xi^2+\eta^2)
                   +\frac{\hbar^2}{2m}(k_1^2-\viert)(\xi^4-1)
                   +\frac{\hbar^2}{2m}(k_2^2-\viert)(\xi^4-1)}
                                            {4(b\xi^2\eta^2-a)}$
          &&$\underline{\hbox{Parabolic}}$      &\cr
&         &&  &&Displaced elliptic                  &\cr
height2pt&\omit&&\omit&&\omit&&\omit&&\omit&\cr
\noalign{\hrule}
height2pt&\omit&&\omit&&\omit&&\omit&&\omit&\cr
&$V_4$    &&$R_1=X_1+\frac{\hbar^2v_0^2}{m}\frac{v}{bu^2-a}$
              &&$\underline{\hbox{$(u,v)$-System}}$ &\cr
&         &&$R_2=X_2+\frac{\hbar^2v_0^2}{2m}\frac{u^2+v^2}{bu^2-a}$
              &&$\underline{\hbox{Polar}}$          &\cr
&         &&$R_3=K=p_v$
              &&$\underline{\hbox{Parabolic}}$      &\cr
&         &&  &&$\underline{\hbox{Elliptic}}$       &\cr
height2pt&\omit&&\omit&&\omit&&\omit&&\omit&\cr
\noalign{\hrule}}}\end{array}\nonumber\end{eqnarray}
\end{table}
\hfuzz=5pt

\subsection{The Superintegrable Potential $V_1$ on $\DII$}
\message{The Superintegrable Potential V_1 on D_II}
We state the potential $V_1$ in the respective coordinate systems:
\begin{eqnarray}
V_1(u,v)&=&\frac{bu^2-a}{u^2}
\left[\frac{m}{2}\omega^2(u^2+4v^2)+k_1v+\frac{\hbar^2}{2m}
\frac{k_2^2-\viert}{u^2}\right]\enspace,
\\      &=&
\frac{bu^2-a}{u^2}\frac{1}{\xi^2+\eta^2}\left[\frac{m}{2}\omega^2(\xi^6+\eta^6)
-\frac{k_1}{2}(\xi^4-\eta^4)-\hbar^2\frac{k_1^2-\viert}{2m}
\bigg(\frac{1}{\xi^2}+\frac{1}{\eta^2}\bigg)\right]\enspace.\qquad
\label{V1DIIparabolic}
\end{eqnarray}
In flat space, the corresponding potential 
is known as the Holt-potential \cite{GROPOa}. It consists of a radial harmonic
oscillator in one variable (here in the variable $u$), and a harmonic
oscillator plus a linear term in the second variable (here in the
variable $v$). There is an analogue of this potential on the
two-dimensional hyperboloid \cite{GROPOc}, which separates in
horicyclic and semi-circular parabolic coordinates, the limiting cases
of the $(u,v)$-system and the parabolic coordinates, respectively.

\subsubsection{Separation of $V_1$ in the $(u,v)$-System.}
We start with the  $(u,v)$-coordinate system. We formulate the
classical Lagrangian and Hamiltonian, respectively:
\begin{eqnarray}
\CL(u,\dot u,v,\dot v)&=&\frac{m}{2}\frac{bu^2-a}{u^2}
                         (\dot u^2+\dot v^2)-V(u,v)\enspace,
\\
\CH(u,p_u,v,p_v)&=& \frac{1}{2m}\frac{u^2}{bu^2-a}(p_u^2+p_v^2)+V(u,v)\enspace.
\end{eqnarray}
The canonical momenta are
\begin{equation}
p_u=\hi\bigg(\frac{\partial}{\partial u}
+\frac{bu}{bu^2-a}-\frac{1}{u}\bigg),\quad
p_v=\hi\frac{\partial}{\partial v}\enspace.
\end{equation}
The quantum Hamiltonian has the form
\begin{eqnarray}
H&=&-\frac{\hbar^2}{2m}\frac{u^2}{bu^2-a}
\bigg(\frac{\partial^2}{\partial u^2}+\frac{\partial^2}{\partial v^2}\bigg)+V(u,v)
\\
&=&\frac{1}{2m}\frac{u}{\sqrt{bu^2-a}}(p_u^2+p_v^2)\frac{u}{\sqrt{bu^2-a}}+V(u,v)
\enspace.
\end{eqnarray}
Therefore the path integral for $V_1$ in the $(u,v)$-system has the 
following form
\begin{eqnarray}
&&K^{(V_1)}(u'',u',v'',v';T)=\pathint{u}\pathint{v}\frac{bu^2-a}{u^2}
\nonumber\\  &&\qquad\times
\exp\left(\ih\int_0^T\left\{\frac{m}{2f}(\dot u^2+\dot v^2)
-f\left[\frac{m}{2}\omega^2(u^2+4v^2)+k_1v+\frac{\hbar^2}{2m}
\frac{k_2^2-\viert}{u^2}\right]\right\}\dt\right).\qquad\qquad
\label{KDIIV1}
\end{eqnarray}
and we have abbreviated $f=u^2/(bu^2-a)$.
First we separate the $v$-path integration according to
\begin{eqnarray}
&&\pathint{v}\exp\left\{\ih\int_0^T\left[\frac{m}{2}\dot v^2
-\bigg(\frac{m}{2}\omega^2v^2+k_1v\bigg)\right]\dt\right\}
\nonumber\\   &&\qquad
=\sum_{n=0}^\infty\sqrt{\frac{2m\omega}{\pi\hbar}}\,\frac{1}{2^nn!}
\exp\bigg[-\frac{m\omega}{\hbar}({\tilde v}^{\prime\,2}
+{\tilde v}^{\prime\prime\,2})\bigg]
H_n\left(\sqrt{\frac{2m\omega}{\hbar}}\,{\tilde v'}\right)
H_n\left(\sqrt{\frac{2m\omega}{\hbar}}\,{{\tilde v}''}\right)
\e^{-\i E_nT/\hbar},
\nonumber\\   &&
         \\   &&
E_n=\hbar\omega(n+\bhalf)+\frac{k_1^2}{8m\omega^2}\enspace,
\end{eqnarray}
with $\tilde v=v+k_1/4m\omega$, which is the solution for the shifted
harmonic oscillator. Writing for short the wave-functions of the
shifted harmonic oscillator by $\Psi_n^{(HO)}$, we thus obtain:
\begin{eqnarray}
&&K^{(V_1)}(u'',u',v'',v';T)=\sum_{n=0}^\infty
\Psi_n^{(HO)}(\tilde v')\Psi_n^{(HO)}(\tilde v'')
K_n^{(V_1)}(u'',u';T)\qquad\qquad
        \\   &&
K_n^{(V_1)}(u'',u';T)=
\big[f(u')f(u'')]^{-1/4}
\pathint{u}\sqrt{\frac{bu^2-a}{u^2}}
\nonumber\\  &&\qquad\times
\exp\left\{\ih\int_0^T\left[\frac{m}{2f}\dot u^2
-f\bigg(\frac{m}{2}\omega^2u^2+\frac{\hbar^2}{2m}\frac{k_2^2-\viert}{u^2}
\bigg)+E_n\right]\dt\right\}.
\end{eqnarray}
We obtain in the usual way by means of a space-time transformation
\begin{equation}
G_n^{(V_1)}(u'',u';E)=\int_0^\infty K_n^{(V_1)}(u'',u';s'')
\exp\bigg[\ih(bE-E_n)s''\bigg]
\end{equation}
with the transformed path integral given by
\begin{eqnarray}
&&K_n^{(V_1)}(u'',u';s'')
\nonumber\\   &&
=\pathints{u}\exp\left\{\ih\ints \left[\frac{m}{2}(\dot u^2-\omega^2u^2)
-\frac{\hbar^2}{2m}\frac{k_2^2+2maE/\hbar^2-\viert}{u^2}\right]\d s\right\}
\label{KNDIIV1s}
         \\   &&
=\frac{m\omega\sqrt{u'u''}}{\i\hbar\sin\omega s''}
\exp\bigg[-\frac{m\omega}{2\i\hbar}({u'}^2+{u''}^2)\cot\omega s''\bigg]
I_\lambda\bigg(\frac{m\omega u'u''}{\i\hbar\sin\omega s''}\bigg)
\nonumber\\   &&
=\sum_{l=0}^\infty\Psi_l^{(RHO,\lambda)}(u')\Psi_l^{(RHO,\lambda)}(u'')\,
\e^{\i s''\omega(2l+\lambda+1)}\enspace.
\end{eqnarray}
Alternatively we have for the Green function
($\lambda^2=k_2^2+2maE/\hbar^2$)
\begin{eqnarray}
G_n^{(V_1)}(u'',u';E)
=\frac{\Gamma\Big[\bhalf\Big(1+\lambda-\frac{1}{\hbar\omega}(bE-E_n)\Big)\Big]}
{\hbar\omega\sqrt{u'u''}\,\Gamma(1+\lambda)}
W_{\frac{bE-E_n}{2\hbar\omega},\frac{\lambda}{2}}\bigg(\frac{m\omega}{\hbar}
u_>^2\bigg)
M_{\frac{bE-E_n}{2\hbar\omega},\frac{\lambda}{2}}\bigg(\frac{m\omega}{\hbar}
u_<^2\bigg)\,.\quad
\label{GDIIV1}
\end{eqnarray}
The $W_{\mu,\nu}(z)$ are Whittaker functions \cite{GRA}.
We can either evaluate the $s''$-integration or analyze the poles of
the Green function. The latter give the poles in terms of the poles of
the $\Gamma$-function yielding the quantization condition for the
bound states $E_{ln}$:
\begin{equation}
\half\bigg(1+\lambda+\frac{E_n-bE_{ln}}{\hbar\omega}\bigg)=-l\enspace,
\end{equation}
which is equivalent to
\begin{equation}
\hbar\omega\left(2l+n+\frac{3}{2}
+\sqrt{k_2^2+\frac{2maE_{ln}}{\hbar^2}}\,\right)
+\frac{k_1^2}{8m\omega^2}-bE_{ln}=0\enspace.
\end{equation}
Let us analyze this equation in more detail. We obtain similar equations for 
the other potentials, and the present case serves as a standard example for 
those which come later. Let us note that the specific form of the discrete
spectrum and the corresponding wave-functions depend on the special choice
of the parameters $a$ and $b$ and the special space of revolution one
considers. For instance, the plus- respectively the minus-sign in the
square-root expression below may be allowed giving positive normed states
for some cases, and for others the minus sign may be allowed.
Similarly, the radicand of the square-root can be become negative and we may
obtain semi-bound states.

\noindent
The quadratic equation in $E_{ln}$ gives ($\epsilon_{ln}=(2l+n+3/2)$)
\begin{eqnarray}
E_{ln}&=&
\frac{\hbar\omega\epsilon_{ln}}{b}+\frac{k_1^2}{8mb\omega^2}
+\frac{am\omega^2}{b}\pm\frac{1}{b^2}
\sqrt{a^2m^2\omega^4+b^2\omega^2\hbar^2k_2^2+2abm\hbar\omega^3\epsilon_{ln}
+\frac{ab}{4}k_1^2},
\nonumber\\   &
\label{ElnDIIV1}
\\    (l,n\to\infty)&\simeq&
\frac{\hbar\omega}{b}(2l+n+\hbox{$\frac{3}{2}$})+\frac{k_1^2}{8mb\omega^2}
+\frac{a}{b}m\omega^2+O(\sqrt{\epsilon_{ln}})\enspace,
\\    \hbox{($a\!=\!-1,\!b\!=\!1$)}&=&
\hbar\omega\epsilon_{ln}+\frac{k_1^2}{8m\omega^2}-m\omega^2
\pm\sqrt{m^2\omega^4+\omega^2\hbar^2k_2^2-2m\hbar\omega^3\epsilon_{ln}
-\frac{k_1^2}{4}}\enspace,
\end{eqnarray}
In the latter (special) case this gives bound states for 
$m^2\omega^4+\hbar^2\omega^2\hbar^2k_2^2-2m\hbar
\omega^3\epsilon_{ln}-k_1^2/4\geq0$, i.e., the number of levels is determined by
\begin{equation}
2l+n\leq
\frac{\hbar k_2^2}{2m\omega}+\frac{m\omega}{2\hbar}
-\frac{k_1^2}{8m\hbar\omega^3}-\frac{3}{2}\enspace,
\end{equation}
otherwise we may have semi-bound states, that is 
bound  states with energy $\Re(E_{ln})$ and with a decay width
$\Im(E_{ln})$. They are located in the continuous spectrum.
In particular, we have a ground state
\begin{eqnarray}
E_{00}&=&
\frac{3\hbar\omega}{2b}+\frac{k_1^2}{8mb\omega^2}
+\frac{am\omega^2}{b}\pm\frac{1}{b^2}
\sqrt{a^2m^2\omega^4+b^2\omega^2\hbar^2k_2^2+3abm\hbar\omega^3+\frac{ab}{4}k_1^2},
\end{eqnarray}
Note that if the radicand of the square root equals
the upper bound of the energy-levels for the case $ab<1$ we get:
\begin{eqnarray}
E_{\rm upper-bound}&=&
\frac{b\hbar^2k_2^2}{2|ab|m}+\frac{ma\omega^2}{|ab|}\bigg(\frac{1}{2b}-1\bigg),
\\ 
&=&\frac{\hbar^2k_2^2}{2m}-\frac{m}{2}\omega^2\qquad(a=-1,b=1)\enspace.
\end{eqnarray}
The spectrum is similar to the spectrum of the Holt potential:
Flat Euclidean space corresponds to $a=0$, then (\ref{ElnDIIV1}) is identical
with the result of \cite{GROPOa}.

Note that different energy spectra emerge depending on the signs of the 
parameters $a$ and $b$. For both parameters positive the discrete spectrum 
cannot be simultaneously located in the continuous spectrum. For $b$ negative,
the properties of the space $\DII$ must be further analyzed, if a discrete 
spectrum with negative infinite values is allowed (which is the case for the 
single-sheeted hyperboloid). 

In order to extract the continuous spectrum we consider the
dispersion relation \cite{GROb}
\begin{equation}
I_\lambda (z)=\frac{2}{\pi^2}\int_0^\infty
\frac{\d p\,\sinh\pi p}{p^2-\lambda^2}K_{\i p}(z)\enspace.
\label{I-lambda-to-K}
\end{equation}
This gives
\begin{eqnarray}
&&G_n^{(V_1)}(u'',u';E)
=\sqrt{u'u''}\int_0^\infty\frac{\omega\d s''}{\i\hbar\sin\omega s''}
\nonumber\\   &&\qquad\qquad\times
\exp\bigg[\ih s''(bE-E_N)-\frac{m\omega}{2\i\hbar}
({u'}^2+{u''}^2)\cot\omega s''\bigg]
I_\lambda\bigg(\frac{m\omega u'u''}{\i\hbar\sin\omega s''}\bigg)
\nonumber\\   &&
=\frac{\hbar^2}{\pi^2}\frac{1}{2m\omega\sqrt{u'u''}}
\int_0^\infty\frac{\d p\,\sinh\pi p}{\frac{\hbar^2}{2m|a|}(p^2+k_2^2)-E}
\nonumber\\   &&\qquad\qquad\times
\Big|\Gamma\Big[\bhalf(1+\i p-\frac{bE-E_n}{\hbar\omega})\Big]\Big|^2
W_{\frac{bE-E_n}{2\hbar\omega},\frac{\i p}{2}}
\bigg(\frac{m\omega}{\hbar}{u''}^2\bigg)
W_{\frac{bE-E_n}{2\hbar\omega},\frac{\i p}{2}}
\bigg(\frac{m\omega}{\hbar}{u'}^2\bigg)\enspace.\qquad\qquad
\end{eqnarray}
The continuous spectrum has the form
\begin{equation}
E_p=\frac{\hbar^2}{2m|a|}(p^2+k_2^2)\enspace,
\end{equation}
and the wave-functions are
\begin{eqnarray}
\Psi_{pn}(u)=\frac{\hbar}{\pi}
\sqrt{\frac{p\sinh\pi p}{2m\omega u}}
\Gamma\bigg[\half\bigg(1+\i p-\frac{bE-E_n}{\hbar\omega}\bigg)\bigg]
W_{\frac{bE-E_n}{2\hbar\omega},\frac{\i p}{2}}
\bigg(\frac{m\omega}{\hbar}u^2\bigg)\enspace.
\end{eqnarray}
Note that for $k_2=\pm\half$, i.e. the radial potential equals zero,
we obtain the case from the free motion on $\DII$.

Finally we state the kernel $K^{(V_1)}(T)$ and the Green function
$G^{(V_1)}(E)$ which have the form
\begin{eqnarray}
&&
\!\!\!\!\!\!\!\!
K^{(V_1)}(u'',u',v'',v';T)
=\sum_{n=0}^\infty
\Psi_n^{(HO)}(\tilde v')\Psi_n^{(HO)}(\tilde v'')
\nonumber\\   &&\!\!\!\!\!\!\!\!
\qquad\times
\Bigg\{\sum_{l=0}^\infty N_{ln}^2\Psi_l^{(RHO,\lambda)}(u')\Psi_l^{(RHO,\lambda)}(u'')\,
\e^{-\i TE_{ln}/\hbar}
+\int_0^\infty\d p\Psi_{pn}^*(u'')\Psi_{pn}(u')\,\e^{-\i TE_p/\hbar}\Bigg\}\qquad\qquad
         \\   &&
\!\!\!\!\!\!\!\!
G^{(V_1)}(u'',u',v'',v';E)
=\sum_{n,l=0}^\infty 
\Psi_n^{(HO)}(\tilde v')\Psi_n^{(HO)}(\tilde v'')
\nonumber\\   &&\!\!\!\!\!\!\!\!
\qquad\times
\frac{\Gamma\Big[\bhalf\Big(1+\lambda-\frac{1}{\hbar\omega}(bE-E_n)\Big)\Big]}
{\hbar\omega\sqrt{u'u''}\,\Gamma(1+\lambda)}
W_{\frac{bE-E_n}{2\hbar\omega},\frac{\lambda}{2}}\bigg(\frac{m\omega}{\hbar}
u_>^2\bigg)
M_{\frac{bE-E_n}{2\hbar\omega},\frac{\lambda}{2}}\bigg(\frac{m\omega}{\hbar}
u_<^2\bigg)\,.\qquad
\end{eqnarray}
The normalization constant $N_{ln}$ emerges form evaluating the residuum of the Green function
(\ref{GDIIV1}) at the energy $E_{ln}$ as given in (\ref{ElnDIIV1}).

\subsubsection{Separation of $V_1$ in Parabolic Coordinates on $\DII$}
The classical Lagrangian and Hamiltonian are given by
\begin{eqnarray}
\CL(\xi,\dot\xi,\eta,\dot\eta)&=&\frac{m}{2}\frac{b\xi^2\eta^2-a}{\xi^2\eta^2}
(\xi^2+\eta^2)(\dot\xi^2+\dot\eta^2)-V(\xi,\eta)\enspace,
\\
\CH(\xi,p_\xi,\eta,p_\eta)&=&\frac{m}{2}\frac{\xi^2\eta^2}{b\xi^2\eta^2-a}
\frac{p_\xi^2+p_\eta^2}{\xi^2+\eta^2}+V(\xi,\eta)\enspace.
\end{eqnarray}
The canonical momenta are given by
\begin{eqnarray}
p_\xi&=&\hi\bigg(\frac{\partial}{\partial\xi}
+\frac{b\xi+a/\xi^3}{\sqrt{g}}\bigg)\enspace,
\\
p_\eta&=&\hi\bigg(\frac{\partial}{\partial\eta}
+\frac{b\eta+a/\eta^3}{\sqrt{g}}\bigg)\enspace.
\end{eqnarray}
The quantum Hamiltonian has the form:
\begin{eqnarray}
H&=&-\frac{\hbar^2}{2m}
\bigg(b\xi^2+b\eta^2-\frac{a}{\xi^2}-\frac{a}{\eta^2}\bigg)^{-1}
\bigg(\frac{\partial^2}{\partial\xi^2}+\frac{\partial^2}{\partial\eta^2}\bigg)
+V(\xi,\eta)\\
&=&\frac{1}{2m}
\bigg(b\xi^2+b\eta^2-\frac{a}{\xi^2}-\frac{a}{\eta^2}\bigg)^{-1/2}
(p_\xi^2+p_\eta^2)
\bigg(b\xi^2+b\eta^2-\frac{a}{\xi^2}-\frac{a}{\eta^2}\bigg)^{-1/2}+V(\xi,\eta)
\,.\qquad
\end{eqnarray}
We obtain for the path integral in parabolic coordinates, c.f.
(\ref{V1DIIparabolic}),
($1/f(\xi,\eta)=(b\xi^2\eta^2-a)/\xi^2\eta^2$):
\begin{eqnarray}
&&\!\!\!\!K^{(V_1)}(\xi'',\xi',\eta'',\eta';T)
=\pathint{\xi}\pathint{\eta}
\frac{b\xi^2\eta^2-a}{\xi^2\eta^2}(\xi^2+\eta^2)
\nonumber\\  &&\!\!\!\!\times
\exp\left\{\ih\int_0^T\left[\frac{m}{2f}(\xi^2+\eta^2)
(\dot\xi^2+\dot\eta^2)
-f\bigg(\frac{m}{2}\omega^2(u^2+4v^2)+
k_1v+\frac{\hbar^2}{2m}
\frac{k_2^2-\viert}{u^2}\bigg)\right]\dt\right\}.\qquad
\end{eqnarray}
Performing the space-time transformation yields
\begin{equation}
G^{(V_1)}(\xi'',\xi',\eta'',\eta';E)
=\int_0^\infty \d s''K^{(V_1)}(\xi'',\xi',\eta'',\eta';s'')
\end{equation}
with the transformed path integral given by
\begin{eqnarray}
&&\!\!\!\!\!\!\!\!\!\!\!\!
K^{(V_1)}(\xi'',\xi',\eta'',\eta';s'')
\nonumber\\  &&\!\!\!\!\!\!\!\!\!\!\!\!
=\pathints{\xi}
\exp\left[\ih\ints \left(\frac{m}{2}(\dot\xi^2-\omega^2\xi^6)
-\frac{k_1}{2}\xi^4
-Eb\xi^2-\frac{\hbar^2}{2m}\frac{k_1^2+2maE/\hbar^2-\viert}{\xi^2}
\right)\d s\right]
\nonumber\\  &&\!\!\!\!\!\!\!\!\!\!\!\!\quad\times
\pathints{\eta}\exp\left[\ih\ints \left(\frac{m}{2}
(\dot\eta^2-\omega^2\eta^6)+\frac{k_1}{2}\eta^4
-Eb\eta^2-\frac{\hbar^2}{2m}
\frac{k_1^2+2maE/\hbar^2-\viert}{\eta^2}\right)\d s\right].
\nonumber\\  &&\!\!\!\!\!\!\!\!\!\!\!\!
\end{eqnarray}
These path integrals are due to the anharmonic terms in $\xi$ and $\eta$
not tractable, a well-known fact due to its relation to the Holt-potential.

\subsection{The Superintegrable Potential $V_2$ on $\DII$.}
\message{The Superintegrable Potential V_2 on D_II.}
We consider the potential $V_2$. The corresponding quantum mechanical
problem is separable in the $(u,v)$-system, in polar and elliptic
coordinates. 
First we state the potential $V_2$ in the respective coordinate systems:
\begin{eqnarray}
V_2(u,v)&=&\frac{u^2}{bu^2-a}\left[\frac{m}{2}\omega^2(u^2+v^2)
+\frac{\hbar^2}{2m}\bigg(\frac{k_1^2-\viert}{u^2}
+\frac{k_2^2-\viert}{v^2}\bigg)\right]\enspace,
        \\    &=&
\frac{\vrho^2\cos^2\vtheta}{b\vrho^2\cos^2\vtheta-a}
\left[\frac{m}{2}\omega^2\vrho^2
+\frac{\hbar^2}{2m\vrho^2}\bigg(\frac{k_1^2-\viert}{\cos^2\vtheta}
+\frac{k_2^2-\viert}{\sin^2\vtheta}\bigg)\right]\enspace,
         \\   &=&
\frac{f}{\cosh^2\omega-\cos^2\vphi}
\Bigg[\frac{m}{2}d^2\omega^2(\cosh^2\omega\sinh^2\omega+\sin^2\vphi\cos^2\vphi)
\qquad\qquad\qquad
\nonumber\\   &&\qquad\qquad\qquad
+\frac{\hbar^2}{2md^2}\Bigg(
 \frac{k_1^2-\viert}{\cos^2\vphi}+\frac{k_2^2-\viert}{\sin^2\vphi}
-\frac{k_1^2-\viert}{\cosh^2\omega}+\frac{k_2^2-\viert}{\sinh^2\omega}
\Bigg)\Bigg]\enspace.
\end{eqnarray}
The potential $V_2$ can be interpreted as a two-dimensional oscillator 
with radial term similarly as its analogue in flat space.
Note that a Higgs-like harmonic oscillator on $\DII$ could have a form
according to (with the limiting case the Higgs-oscillator on the hyperboloid)
\begin{eqnarray}
V_{\rm Higgs}&=&\frac{m}{2}\omega^2\frac{u^2}{bu^2-a}
\left(1-\frac{4u^2}{(1+u^2+v^2)^2}\right)
=\frac{m}{2}\omega^2\frac{\vrho^2\cos^2\vtheta}{b\vrho^2\cos^2\vtheta-a}
\left(1-\frac{4\vrho^2\cos^2\vtheta}{(1+\vrho^2)^2}\right)
\nonumber\\   &=&
\frac{m}{2}\omega^2\bigg(\frac{b\,\e^{2\tau_2}}{\cosh^2\tau_1}-a\bigg)^{-1}
\left(1-\frac{1}{\cosh^2\tau_1\cosh^2\tau_2}\right)\enspace,
\end{eqnarray}
with $\vrho=e^{\tau_2}$, $\cos\vtheta=1/\cosh\tau_1$, $\tau_{1,2}$
being equidistant coordinates. The corresponding path integral cannot
be solved, and $V_{\rm Higgs}$ is not superintegrable in $\DII$ either. 

\subsubsection{Separation of $V_2$ in the $(u,v)$-System.}
We start with the consideration in the $(u,v)$-system, and
the path integral has the form
\begin{eqnarray}
&&K^{(V_2)}(u'',u',v'',v';T)
=\pathint{u}\pathint{v}\frac{bu^2-a}{u^2}
\nonumber\\  &&\qquad\times
\exp\left\{\ih\int_0^T\left[\frac{m}{2f}(\dot u^2+\dot v^2)
-f\frac{m}{2}\omega^2(u^2+v^2)-f\frac{\hbar^2}{2m}\bigg(
\frac{k_1^2-\viert}{u^2}+\frac{k_2^2-\viert}{v^2}\bigg)
\right]\dt\right\}\qquad
\label{KDIIV2}
         \\  &&
=\sum_{n=0}^\infty\Psi_n^{(RHO,k_2)}(v'')\Psi_n^{(RHO,k_2)}(v')
\big[f(u')f(u'')]^{-1/4}
\pathint{u}\sqrt{\frac{bu^2-a}{u^2}}
\nonumber\\  &&\qquad\times
\exp\left\{\ih\int_0^T\left[\frac{m}{2f}\dot u^2
-f\bigg(\frac{m}{2}u^2+\frac{\hbar^2}{2m}\frac{k_1^2-\viert}{u^2}
+E_n\bigg)\right]\dt\right\},
\end{eqnarray}
where $E_n=\hbar\omega(2n+|k_2|+1)$. Performing a space-time
transformation in the usual way yields:
\begin{equation}
G_n^{(V_2)}(u'',u';E)=
\int_0^\infty \d s''\e^{\i s''(bE-E_n)/\hbar}K_n^{(V_2)}(u'',u';s'')
\end{equation}
with the transformed path integral given by
\begin{equation}
K_n^{(V_2)}(u'',u';s'')=
\pathints{u}\exp\left\{\ih\ints \left[
\frac{m}{2}(\dot u^2-\omega^2u^2)
-\frac{\hbar^2}{2m}\frac{\lambda_1^2-\viert}{u^2}\right]\d s\right\},
\end{equation}
where $\lambda_1^2=k_1^2+2maE/\hbar^2$.
This path integral has almost the same form as the path integral
(\ref{KNDIIV1s}), the only difference being another $E_n$. Thus we can
write the solution as follows:
\begin{eqnarray}
K_n^{(V_2)}(u'',u';s'')
&=&\frac{m\omega\sqrt{u'u''}}{\i\hbar\sin\omega s''}
\exp\bigg[-\frac{m\omega}{2\i\hbar}({u'}^2+{u''}^2)\cot\omega
  s''\bigg]
I_{\lambda_1}\bigg(\frac{m\omega u'u''}{\i\hbar\sin\omega s''}\bigg)
\nonumber\\  
&=&\sum_{l=0}^\infty\Psi_l^{(RHO,\lambda_1)}(u')\Psi_l^{(RHO,\lambda_1)}(u'')\,
\e^{\i s''\omega(2l+\lambda_1+1)]}\enspace,
\end{eqnarray}
and alternatively we have for the Green function
\begin{eqnarray}
G_n^{(V_2)}(u'',u';E)\!=\!\frac{\Gamma\Big[\bhalf
\Big(1+\lambda_1-\frac{1}{\hbar\omega}(bE-E_n)\Big)\Big]}
{\hbar\omega\sqrt{u'u''}\,\Gamma(1+\lambda_1)}
W_{\frac{bE-E_n}{2\hbar\omega},\frac{\lambda_1}{2}}\bigg(\frac{m\omega}{\hbar}
u_>^2\bigg)
M_{\frac{bE-E_n}{2\hbar\omega},\frac{\lambda_1}{2}}\bigg(\frac{m\omega}{\hbar}
u_<^2\bigg).\quad
\label{GDIIV2}
\end{eqnarray}
We can either evaluate the $s''$-integration or analyze the poles of
the Green function. The latter give the poles in terms of the poles of
the $\Gamma$-function yielding the quantization condition for the
bound states $E_{ln}$:
\begin{equation}
\half\bigg(1+\lambda_1+\frac{E_n-bE_{ln}}{\hbar\omega}\bigg)=-l\enspace,
\end{equation}
which is equivalent to
\begin{equation}
\hbar\omega\left(2l+2n+2+|k_2|+\sqrt{k_1^2+\frac{2maE_{ln}}{\hbar}}\,\right)
-bE_{ln}=0\enspace.
\label{ElnDIIV20}\end{equation}
The quadratic equation in $E_{ln}$ gives
($\epsilon_{ln}=(2l+2n+2+|k_2|)$)
\begin{eqnarray}
E_{ln}&=&
\frac{\hbar\omega\epsilon_{ln}}{b}+\frac{a}{b}m\omega^2
-\frac{1}{b^2}\sqrt{a^2m^2\omega^4+b^2\hbar^2\omega^2k_1^2
+2abm\hbar\omega^3\epsilon_{ln}}\,,
\label{ElnDIIV2}
\\
\hbox{($a\!=\!-1,\! b\!=\!1$)} &=&
\hbar\omega\epsilon_{ln}-m\omega^2-
\sqrt{m^2\omega^4+\hbar^2\omega^2k_1^2-2m\omega^2\epsilon_{ln}}\enspace,
\\   (l,n\to\infty)&\simeq&
\hbar\omega\epsilon_{ln}-m\omega^2\enspace.
\end{eqnarray}
This gives for the special case bound states for 
$m^2\omega^4+\hbar^2\omega^2k_1^2-2m\omega^2\epsilon_{ln}\geq0$, 
otherwise we can infer for semi-bound states, that is 
bound  states with energy $\Re(E_{ln})$ and with a decay width
$\Im(E_{ln})$. They are located in the continuous spectrum.
Again, the limiting case of flat space emerges from $a=0,b=1$
\begin{equation}
E_{ln}=\hbar\omega(2l+2n+|k_1|+|k_2|+2)\enspace.
\end{equation}
Finally we state the kernel $K^{(V_2)}(T)$ and the Green function
$G^{(V_2)}(E)$ which have the form
\begin{eqnarray}
&&K^{(V_2)}_{\hbox{descrete}}(u'',u',v'',v';T)
\nonumber\\   &&
=\sum_{n,l=0}^\infty N_{ln}^2
\Psi_n^{(RHO,k_2)}(v')\Psi_n^{(RHO,k_2)}(v'')
\Psi_l^{(RHO,\lambda_1)}(u')\Psi_l^{(RHO,\lambda_1)}(u'')\,
\e^{-\i TE_{ln}/\hbar},
         \\   &&
G^{(V_2)}(u'',u',v'',v';E)
=\sum_{n=0}^\infty 
\Psi_n^{(RHO,k_2)}(v')\Psi_n^{(RHO,k_2)}(v'')
\nonumber\\   &&\qquad\qquad\times\frac{\Gamma\Big[
\bhalf\Big(1+\lambda_1-\frac{1}{\hbar\omega}(bE-E_n)\Big)\Big]}
{\hbar\omega\sqrt{u'u''}\,\Gamma(1+\lambda_1)}
W_{\frac{bE-E_n}{2\hbar\omega},\frac{\lambda_1}{2}}\bigg(\frac{m\omega}{\hbar}
u_>^2\bigg)
M_{\frac{bE-E_n}{2\hbar\omega},\frac{\lambda_1}{2}}\bigg(\frac{m\omega}{\hbar}
u_<^2\bigg)\,.\quad\qquad
\end{eqnarray}
The normalization constant $N_{ln}$ emerges form evaluating the residuum of
the Green function 
(\ref{GDIIV2}) at the energy $E_{ln}$ as given in (\ref{ElnDIIV2}).
We omit the continuous part of $K^{(V_2)}$ due to its similarity to 
the case of $V_1$.

\subsubsection{Separation of $V_2$ in Polar Coordinates.}
The potential $V_2$ is also separable in polar coordinates on $\DII$.
In polar coordinates the classical Lagrangian and Hamiltonian are given by
\begin{eqnarray}
\CL(r,\dot r,\vtheta,\dot\vtheta)&=&\frac{m}{2}
  \bigg(b-\frac{a}{\vrho^2\cos^2\vtheta}\bigg)
(\dot \vrho^2+\vrho^2\dot\vtheta^2)-V(\vrho,\vtheta)\enspace,
\\
\CH(\vrho,p_\vrho,\vtheta,p_\vtheta)&=&\frac{1}{2m}
\bigg(b-\frac{a}{\vrho^2\cos^2\vtheta}\bigg)^{-1}
\bigg(p_\vrho^2+\frac{1}{\vrho^2}p_\vtheta^2\bigg)+V(\vrho,\vtheta)\enspace.
\end{eqnarray}
The momentum operators are
\begin{eqnarray}
p_\vrho&=&\hi\bigg[\frac{\partial}{\partial\vrho}
+\bigg(\frac{b\vrho\cos^2\vtheta}{b\cos^2\vtheta\vrho^2-a}-\frac{1}{2\vrho}
\bigg)\bigg]\enspace,
\\
p_\vtheta&=&\hi\bigg[\frac{\partial}{\partial\vtheta}+\bigg(\tan\vtheta
-\frac{b\vrho^2\sin\vtheta\cos\vtheta}{b\vrho^2\cos^2\vtheta-a}\bigg)\bigg]
\enspace,
\end{eqnarray}
and the quantum Hamiltonian is given by:
\begin{eqnarray}
H&=&-\frac{\hbar^2}{2m}\bigg(b-\frac{a}{\vrho^2\cos^2\vtheta}\bigg)^{-1}
 \bigg(\frac{\partial^2}{\partial \vrho^2}
  +\frac{1}{\vrho}\frac{\partial}{\partial\vrho}
  +\frac{1}{\vrho^2}\frac{\partial^2}{\partial\vtheta^2}\bigg)+V(\vrho,\vtheta)
\\
&=&\frac{1}{2m}f^{1/2}
\bigg(p_\vrho^2+\frac{1}{\vrho^2}p_\vtheta^2\bigg)f^{1/2}+V(\vrho,\vtheta)
-f\frac{\hbar^2}{8m\vrho^2}\enspace.
\nonumber\\
\end{eqnarray}
with the abbreviation $1/f=b-a/\vrho^2\cos^2\vtheta$. 
Hence, we get for the path integral 
\begin{eqnarray}
&&K^{(V_2)}(\vrho'',\vrho',\vtheta'',\vtheta';T)=
\pathint{\vrho}\vrho\pathint{\vtheta}
\bigg(b-\frac{a}{\vrho^2\cos^2\vtheta}\bigg)
\nonumber\\  &&\ \times
\exp\left(\ih\int_0^T\left\{
\frac{m}{2f}(\dot\vrho^2+\vrho^2\dot\vtheta^2) 
-f\left[\frac{m}{2}\omega^2\vrho^2+\frac{\hbar^2}{2m\vrho^2}
\bigg(\frac{k_1^2-\viert}{\cos^2\vtheta}+\frac{k_2^2-\viert}{\sin^2\vtheta}
+\viert\bigg)\right]\right\}\dt\right)\,.\qquad\quad
\end{eqnarray}
Performing the space time transformation with the function $f$ yields
\begin{equation}
G^{(V_2)}(\vrho'',\vrho',\vtheta'',\vtheta';E)
=\int_0^\infty\d s''\e^{\i s'' bE/\hbar}
K^{(V_2)}(\vrho'',\vrho',\vtheta'',\vtheta';s'')
\end{equation}
and the transformed path integral given by
($\lambda_1^2=k_1^2+2maE/\hbar^2$)
\begin{eqnarray}
&&K^{(V_2)}(\vrho'',\vrho',\vtheta'',\vtheta';s'')=
\pathints{\vrho}\pathints{\vtheta}\vrho
\nonumber\\  &&\qquad\times
\exp\left\{\ih\ints \left[
\frac{m}{2}(\dot\vrho^2+\vrho^2\dot\vtheta^2) 
-\frac{m}{2}\omega^2\vrho^2-\frac{\hbar^2}{2m\vrho^2}
\bigg(\frac{\lambda_1^2-\viert}{\cos^2\vtheta}
+\frac{k_2^2-\viert}{\sin^2\vtheta}+\viert\bigg)\right]\d s\right\}\qquad
          \\  &&
=\frac{1}{\sqrt{\vrho'\vrho''}}\sum_{n=0}^\infty
\Phi_n^{(\lambda_1,k_2)}(\vtheta'')
\Phi_n^{(\lambda_1,k_2)}(\vtheta')
\nonumber\\  &&\qquad\times
\pathints{\rho}
\exp\left\{\ih\ints \left[\frac{m}{2}(\dot\vrho^2-\omega^2\vrho^2)
-\frac{\hbar^2}{2m}\frac{\lambda_2^2-\viert}{\vrho^2}\right]\d s\right\}
          \\  &&
=\frac{1}{\sqrt{\vrho'\vrho''}}\sum_{n=0}^\infty
\Phi_n^{(\lambda_1,k_2)}(\vtheta'')
\Phi_n^{(\lambda_1,k_2)}(\vtheta')\sum_{l=0}^\infty
\Psi_l^{(RHO,\lambda_2)}(\vrho'')\Psi_l^{(RHO,\lambda_2)}(\vrho')
\,\e^{-\i s''E_l/\hbar}\enspace.
\end{eqnarray}
Here denote $E_l=\hbar\omega(2l+\lambda_2+1)$, and the quantity 
$\lambda_2$ is defined by means of the energy-spectrum of the 
P\"oschl--Teller spectrum
\begin{equation}
\frac{\hbar^2}{2m}(2n+1+\lambda_1+|k_2|)
=\frac{\hbar^2}{2m}\lambda_2^2\enspace.
\end{equation}
The $\Phi_n^{(k_1,k_2)}(\beta)$ are the wave-functions of the 
P\"oschl--Teller potential, which are given by \cite{BJb,DURb,FLM,KLEMUS}
\begin{eqnarray}
  V(x)&=&\hbarm\bigg(
  {\alpha^2-{1\over4}\over\sin^2x}+{\beta^2-{1\over4}\over\cos^2x}\bigg)
           \\  
  \Phi_n^{(\alpha,\beta)}(x)
  &=&\bigg[2(\alpha+\beta+2l+1)
  {l!\Gamma(\alpha+\beta+l+1)\over\Gamma(\alpha+l+1)\Gamma(\beta+l+1)}
  \bigg]^{1/2}
  \nonumber\\   &&\qquad\qquad\times
  (\sin x)^{\alpha+1/2}(\cos x)^{\beta+1/2}
  P_n^{(\alpha,\beta)}(\cos2x)\enspace.
\end{eqnarray}
The $P_n^{(\alpha,\beta)}(z)$ are Gegenbauer polynomials \cite{GRA}.
Performing the $s''$-integration give poles in the Green function for
\begin{equation}
\hbar\omega(2l+2n+2++\lambda_1+|k_2|)-bE_{ln}=0\enspace.
\end{equation}
This is identical to (\ref{ElnDIIV20}), as it should be.
Concerning the discrete spectrum we can state the kernel as follows
\begin{eqnarray}
&&K^{(V_2)}_{\hbox{descrete}}(\vrho'',\vrho',\vtheta'',\vtheta';T)
=\frac{1}{\sqrt{\vrho'\vrho''}}\sum_{n=0}^\infty
\Phi_n^{(\lambda_1,k_2)}(\vtheta'')
\Phi_n^{(\lambda_1,k_2)}(\vtheta')
\nonumber\\  &&\qquad\qquad\qquad\qquad\times
\sum_{l=0}^\infty N_{ln}^2
\Psi_l^{(RHO,\lambda_2)}(\vrho'')\Psi_l^{(RHO,\lambda_2)}(\vrho')\,
\e^{-\i s''E_{ln}/\hbar}\enspace,
\end{eqnarray}
with $N_{ln}$ defined by the residuum of the Green function
at the energy $E_{ln}$ as given in (\ref{ElnDIIV2}).

\subsubsection{Separation of $V_2$ in Elliptic Coordinates on $\DII$}
The free classical Lagrangian and Hamiltonian are given by
\begin{eqnarray}
\CL(\omega,\dot\omega,\vphi,\dot\vphi)&=&
\frac{m}{2}\frac{bd^2\cosh^2\omega\cos^2\vphi-a}{\cosh^2\omega\cos^2\vphi}
        (\cosh^2\omega-\cos^2\vphi)(\dot\omega^2+\dot\vphi^2)
\nonumber\\
&=&\frac{m}{2}\left[
\bigg(bd^2\cosh^2\omega+\frac{a}{\cosh^2\omega}\bigg)
-\bigg(bd^2\cos^2\vphi+\frac{a}{\cos^2\vphi}\bigg)\right]
(\dot\omega^2+\dot\vphi^2)\enspace,\qquad
\\
\CH(\omega,p_\omega,\vphi,p_\vphi)&=&
\frac{1}{2m}\frac{\cosh^2\omega\cos^2\vphi}
{(bd^2\cosh^2\omega\cos^2\vphi-a)(\cosh^2\omega-\cos^2\vphi)}
(p_\omega^2+p_\vphi^2)\enspace.
\end{eqnarray}
In the following we use 
$$\sqrt{g}=\frac{bd^2\cosh^2\omega\cos^2\vphi-a}{\cosh^2\omega\cos^2\vphi}
        (\cosh^2\omega-\cos^2\vphi)\enspace.$$
For the momentum operators we obtain
\begin{eqnarray}
p_\omega&=&\hi\left[\frac{\partial}{\partial\omega}
   +\frac{\tanh\omega}{\sqrt{g}}\bigg(bd^2\cosh^2\omega-\frac{a}{\cosh^2\omega}
\bigg)\right]\enspace,
\\
p_\vphi&=&\hi\left[\frac{\partial}{\partial\vphi}
   +\frac{\tan\vphi}{\sqrt{g}}\bigg(bd^2\cos^2\vphi-\frac{a}{\cos^2\vphi}
\bigg)\right]\enspace.
\end{eqnarray}
This gives for the quantum Hamiltonian
\begin{eqnarray}
H&=&-\frac{\hbar^2}{2m}
\frac{\cosh^2\omega\cos^2\vphi}
{(bd^2\cosh^2\omega\cos^2\vphi-a)(\cosh^2\omega-\cos^2\vphi)}
     \bigg(\frac{\partial^2}{\partial\omega^2}
           +\frac{\partial^2}{\partial\vphi^2}\bigg)
\nonumber\\
&=&\frac{1}{2m}\frac{1}{\sqrt[4]{g}}(p_\omega^2+p_\phi^2)\frac{1}{\sqrt[4]{g}}
\enspace.
\end{eqnarray}
Therefore we obtain for the path integral
($1/f=(bd^2\cosh^2\omega\cos^2\vphi-a)/\cosh^2\omega\cos^2\vphi$)
\begin{eqnarray}
&&\!\!\!\!\!\!\!\!
K^{(V_2)}(\omega'',\omega',\vphi'',\vphi';T)
=\pathint{\omega}\pathint{\vphi}\frac{(\cosh^2\omega-\cos^2\vphi)}{f}
\nonumber\\ &&\!\!\!\!\!\!\!\!\qquad\times
\exp\Bigg\{\ih\int_0^T\Bigg[\frac{m}{2f}
(\cosh^2\omega-\cos^2\vphi)(\dot\omega^2+\dot\vphi^2)
\nonumber\\ &&\!\!\!\!\!\!\!\!\qquad\qquad\qquad\qquad\qquad\qquad\qquad\qquad
-f\frac{m}{2}\omega^2(u^2+v^2)-f\frac{\hbar^2}{2m}
\bigg(\frac{k_1^2-\viert}{u^2}+\frac{k_2^2-\viert}{v^2}\bigg)\Bigg]
\dt\Bigg\}\qquad\qquad
\nonumber\\  &&\!\!\!\!\!\!\!\!
=\int_{-\infty}^\infty\frac{\d E}{2\pi\hbar}\,\e^{-\i ET/\hbar}
\int_0^\infty \d s'' K^{(V_2)}(\omega'',\omega',\vphi'',\vphi';s'')\enspace,
\end{eqnarray}
with the transformed path integral 
$K^{(V_2)}(\omega'',\omega',\vphi'',\vphi';s'')$ given by ($a<0$)
\begin{eqnarray}
&&\!\!\!\!\!\!\!\!
K^{(V_2)}(\omega'',\omega',\vphi'',\vphi';s'')
=\pathints{\omega}
\exp\Bigg\{\ih\ints \Bigg[
\frac{m}{2}(\dot\omega^2-d^2\omega^2\cosh^2\omega\sinh^2\omega)
\nonumber\\ &&\!\!\!\!\!\!\!\!\qquad\qquad\qquad\qquad\qquad\qquad
-\frac{\hbar^2}{2m}\bigg(\frac{-k_1^2-2m|a|E/\hbar^2-\viert}{\cosh^2\omega}
+\frac{k_2^2-\viert}{\sinh^2\omega}\bigg)
+Ebd^2\cosh^2\omega\Bigg]\d s\Bigg\}\qquad
\nonumber\\ &&\!\!\!\!\!\!\!\!\qquad\times
\pathints{\vphi}
\exp\Bigg\{\ih\ints \Bigg[
\frac{m}{2}(\dot\vphi^2-d^2\omega^2\sin^2\vphi\cos^2\vphi)
\nonumber\\ &&\!\!\!\!\!\!\!\!\qquad\qquad\qquad\qquad\qquad\qquad
-\frac{\hbar^2}{2m}\bigg(\frac{k_1^2-2m|a|E/\hbar^2-\viert}{\cos^2\vphi}
+\frac{k_2^2-\viert}{\sin^2\vphi}\bigg)
-Ebd^2\cos^2\vphi\Bigg]\d s\Bigg\}.\qquad
\label{path-ellipticV2}
\end{eqnarray}
We leave these path integrals as they are, because they are not tractable.

\subsection{The Superintegrable Potential $V_3$ on $\DII$.}
\message{The Superintegrable Potential V_3 on D_II.}
We consider the potential $V_3$ and start by expressing $V_3$ in the
respective coordinate systems. We have
\begin{eqnarray}
V_3(u,v)&=&\frac{f}{\sqrt{u^2+v^2}}
\left[-\alpha+\frac{\hbar^2}{2m}\left(\frac{k_1^2-\viert}{\sqrt{u^2+v^2}+v}
+\frac{k_2^2-\viert}{\sqrt{u^2+v^2}-v}\right)\right]\quad
\nonumber\\    
\hbox{(polar coordinates:)\qquad}&=&\frac{2f}{\vrho}
\left[-\alpha+\frac{\hbar^2}{2m\vrho}\left(
\frac{k_1^2-\viert}{1+\sin\vtheta}
+\frac{k_2^2-\viert}{1-\sin\vtheta}\right)\right]
\nonumber\\   
\hbox{(transformation: \qquad\quad}
&&\!\!\!\!\!\!\!\!\!\!\!\!\!\!\!\!\!\!\!\!\!\!\!\!\!\!\!\!\!\!\!\!\!\!\!
\hbox{$\cos\vtheta=\sin2\phi$, $\sin\vtheta=\cos2\phi$,
$\vrho=r^2/2$)}\nonumber\\ 
&=&\frac{f}{r^2}\left[-\alpha+\frac{\hbar^2}{2mr^2}\left(
\frac{k_1^2-\viert}{\cos^2\vphi}
+\frac{k_2^2-\viert}{\cos^2\vphi}\right)\right]
\\   
\hbox{(parabolic coordinates:)}&=&\frac{2f}{\xi^2+\eta^2}
\left[-\alpha+\frac{\hbar^2}{2m}\left(
\frac{k_1^2-\viert}{\xi^2}+\frac{k_2^2-\viert}{\eta^2}\right)\right]
\\ 
\hbox{(rotated elliptic coordinates:)}&=&
\frac{f}{\sqrt{u^2+v^2}}
\frac{1}{\cosh^2\omega'-\cos^2\vphi'}
\Bigg[-{b'}^2\alpha(\cosh^2\omega'-\cos^2\vphi')
\nonumber\\   &&\quad
+\frac{\hbar^2}{2m}\left(
 \frac{k_1^2-\viert}{\cos^2\vphi'}+\frac{k_2^2-\viert}{\sin^2\vphi'}
-\frac{k_1^2-\viert}{\cosh^2\omega'}+\frac{k_2^2-\viert}{\sinh^2\omega'}
\right)\Bigg].\quad
\end{eqnarray}
In the last case the rotated elliptic coordinates are given by
\begin{equation}
u=\frac{{b'}^2}{4}\sinh2\omega'\sin2\vphi'\enspace,\qquad
v=\frac{{b'}^2}{4}(\cosh2\omega'\cos2\vphi'+1)\enspace.
\end{equation}
Due to the complicated structure of the path integral in rotated
elliptic coordinates no closed solution can be stated. We will
omit a path integral discussion of $V_3$ in these coordinates.
The potential $V_3$ can be interpreted as an analogue of the Coulomb
potential. Similarly as in flat space and on the two-dimensional
hyperboloid it is separable in three coordinate systems, i.e. in 
spherical, parabolic, and rotated elliptic coordinates (there are no
conical coordinates in $\DII$). 

\subsubsection{Separation of $V_3$ in Polar Coordinates.}
We start with the investigation of $V_3$ in polar coordinates and we
immediately switch form the $(\rho,\vtheta)$-system to the
$(r,\vphi)$-system. This gives for the path integral:
\begin{eqnarray}
&&\!\!\!\!\!\!\!\!\!\!\!\!
K^{(V_3)}(r'',r',\vphi'',\vphi';T)=\pathint{r}\pathint{\vphi}
\bigg(br^2-\frac{a}{r^2\sin^2\vphi\cos^2\vphi}\bigg)r
\nonumber\\   &&\!\!\!\!\!\!\!\!\!\!\!\!\quad\times
\exp\left(\ih\int_0^T\left\{\frac{m}{2f}(\dot r^2+r^2\dot\vphi^2)
-f\left[-\alpha+\frac{\hbar^2}{2mr^2}\left(
 \frac{k_1^2-\viert}{\cos^2\vphi}
+\frac{k_2^2-\viert}{\cos^2\vphi}-\frac{1}{4}\right)\right]\right\}
\dt\right),\qquad\qquad
\end{eqnarray}
with $1/f=br^2-a/r^2\sin^2\vphi\cos^2\vphi$. Proceeding in the usual
way by means of a space time transformation gives 
\begin{equation}
G^{(V_3)}(r'',r',\vphi'',\vphi';E)
=\int_0^\infty\d s''\e^{\i s''\alpha/\hbar}
K^{(V_3)}(r'',r',\vphi'',\vphi';s'')
\end{equation}
and the path integral $K^{(V_3)}(s'')$ given by
\begin{eqnarray}
&&K^{(V_3)}(r'',r',\vphi'',\vphi';s'')=\pathints{r}\pathints{\vphi}r
\nonumber\\   &&\qquad\times
\exp\left\{\ih\ints \left[\frac{m}{2}(\dot r^2+r^2\dot\vphi^2)
+Ebr^2-\frac{\hbar^2}{2mr^2}
\bigg(\frac{\lambda_2^2-\viert}{\sin^2\vphi}
+\frac{\lambda_1^2-\viert}{\cos^2\vphi}
-\viert\bigg)\right]\d s\right\}
\nonumber\\   &&
=\sum_{n=0}^\infty
\Phi_n^{(\lambda_1,\lambda_2)}(\vphi'')
\Phi_n^{(\lambda_1,\lambda_2)}(\vphi' )
K_n^{(V_3)}(r'',r',s'')\enspace,
\end{eqnarray}
with $\lambda_{1,2}^2=k_{1,2}^2+2maE/\hbar^2$.
The path integral $K_n^{(V_3)}(s'')$ has the form 
\begin{eqnarray}
&&K_n^{(V_3)}(r'',r';s'')=\frac{1}{\sqrt{r'r''}}\pathints{r}
\exp\left[\ih\ints \left(\frac{m}{2}\dot r^2
+Ebr^2-\frac{\hbar^2}{2m}\frac{\Lambda^2-\viert}{r^2}\right)\d s\right]
\nonumber\\   &&
=\frac{m\omega}{\i\hbar\sin\omega s''}
\exp\left[-\frac{m\omega}{2\i\hbar}({r'}^2+{r''}^2)\cot\omega
  s''\right]
I_\Lambda\left(\frac{m\omega r'r''}{\i\hbar\sin\omega s''}\right)
\nonumber\\   &&
=\sqrt{-\frac{m}{2Eb}}\frac{
\Gamma\Big[\half(1+\Lambda-\frac{\alpha}{\hbar}\sqrt{-\frac{m}{2Eb}})\Big]}
{\Gamma(1+\Lambda)\sqrt{r'r''}}
\nonumber\\   &&\qquad\times
M_{\frac{\alpha}{2\hbar}\sqrt{-\frac{m}{2Eb}},\frac{\Lambda}{2}}
\left(\frac{m}{\hbar}\sqrt{-\frac{2Eb}{m}}r_<^2\right)
W_{\frac{\alpha}{2\hbar}\sqrt{-\frac{m}{2Eb}},\frac{\Lambda}{2}}
\left(\frac{m}{\hbar}\sqrt{-\frac{2Eb}{m}}r_>^2\right)
         \\   
&&K_{n,\hbox{discrte}}^{(V_3)}(r'',r';s'')
=\frac{1}{\sqrt{r'r''}}\sum_{l=0}^\infty
\Psi_l^{(RHO,\Lambda)}(r'')\Psi_l^{(RHO,\Lambda)}(r')\,
\e^{-\i\omega(2l+\Lambda+1)s''}\enspace.
\label{DIIV3dis-r-phi}
\end{eqnarray}
This is the usual radial harmonic oscillator solution, and we have set
$\Lambda=2n+\lambda_1+\lambda_2+1$, $\omega^2=-2Eb/m$. The bound
states are determined by the quantization condition
\begin{equation}
2\hbar\omega(l+n+1)+\frac{\hbar^2k_1^2}{2m}+\hbar\omega(\lambda_2+\lambda_3)=0
\enspace,
\end{equation}
or alternatively
\begin{equation}
2(l+n+1)-\frac{\alpha}{\hbar}\sqrt{-\frac{m}{2Eb}}
+\sqrt{k_1^2+\frac{2maE}{\hbar^2}}+\sqrt{k_2^2+\frac{2maE}{\hbar^2}}=0
\enspace.
\label{EnlDIIV3}
\end{equation}
For $a=0$, $b=1$ we recover the two-dimensional flat space Coulomb-spectrum.
In general this is an equation of eighth order in $E$, where no closed
solution can be stated. However, we can study the special
case $k_1=k_2=0$, which gives the quantization condition
(we also take $a<0,b>0$)
\begin{equation}
2(l+n+1)-\frac{\alpha}{\hbar}\sqrt{\frac{m}{2b}}\frac{1}{\sqrt{-E}}
+\frac{2}{\hbar}\sqrt{2m|a|}\sqrt{-E}=0\enspace.
\end{equation}
This is a quadratic equation in $E$ with solution
(only one solution is physical, set $N=l+n+1$)
\begin{eqnarray}
E_{ln}&=&-\frac{\hbar^2N^2}{8m|a|}\left(
\sqrt{1+\frac{2m\alpha}{\hbar^2N^2}\sqrt{\frac{|a|}{b}}}-1\right)^2
         \\   &\simeq&
-\frac{m\alpha^2}{8b\hbar^2N^2}\enspace,\qquad (N\to\infty)\enspace.
\end{eqnarray}
showing a Coulomb-behavior.

In order to extract the continuous spectrum we use
(\ref{I-lambda-to-K}) and obtain for the entire kernel
\begin{eqnarray}
&&\!\!\!\!\!\!\!\!\!\!\!
G^{(V_3)}(r'',r',\vphi'',\vphi';E)
=\frac{\hbar}{\sqrt{r'r''}}
\sum_{n=0}^\infty
\Phi_n^{(\lambda_1,\lambda_2)}(\vphi'')
\Phi_n^{(\lambda_1,\lambda_2)}(\vphi' )
\nonumber\\   &&\!\!\!\!\!\!\!\!\!\!\!\qquad\times
\Bigg\{\sum_{l=0}^\infty\frac{N_{ln}^2}{E_{ln}-E}
\Psi_l^{(RHO,\Lambda)}(r'')\Psi_l^{(RHO,\Lambda)}(r')
+\int_{-\infty}^\infty\frac{1}{E_p-E}
\Psi_{p}^{(RHO,\Lambda)\,*}(r'')\Psi_{p}^{(RHO,\Lambda)}(r')\Bigg\}\qquad
\nonumber\\   &&\!\!\!\!\!\!\!\!\!\!\!
\end{eqnarray}
with the discrete energy spectrum as determined by (\ref{EnlDIIV3}), and
the normalization constant $N_{ln}$ resulting from the residuum in
(\ref{DIIV3dis-r-phi}). The continuous spectrum is given by
($k_<$ denotes the smaller of $k_1,k_2$)
\begin{eqnarray}
\Psi_p^{(RHO)}(r)&=&
\frac{\e^{\pi p/2}}{\sqrt{\pi}}
\frac{\Gamma\Big[\half(1+\Lambda)+\i p\Big]}{\Gamma(1+\Lambda)}
M_{\i p/2,\Lambda/2}\bigg(-\frac{\sqrt{-2mbE}}{\hbar}\,r\bigg)\enspace,
        \\
E_p&=&\frac{\hbar^2}{2m}(p^2+k_<^2)\enspace.
\end{eqnarray}

\subsubsection{Separation of $V_3$ in Parabolic Coordinates.}
Finally we consider $V_3$ in parabolic coordinates. The formulation of
the path integral for a potential on $\DII$ we know already from
$V_1$. We therefore have ($f=b-{a}/{\xi^2\eta^2}$)
\begin{eqnarray}
&&\!\!\!\!\!\!\!\!\!\!\!
K^{(V_3)}(\xi'',\xi',\eta'',\eta';T)=
\pathint{\xi}\pathint{\eta}\bigg(b-\frac{a}{\xi^2\eta^2}\bigg)
(\xi^2+\eta^2)
\nonumber\\   &&\!\!\!\!\!\!\!\!\!\!\!\qquad\times
\exp\left(\ih\int_0^\infty\left\{
\frac{m}{2}f(\xi^2+\eta^2)(\dot\xi^2+\dot\eta^2)
-\frac{1}{f\cdot(\xi^2+\eta^2)}\left[-\alpha
+\frac{\hbar^2}{2m}\bigg(\frac{k_1^2-\viert}{\xi^2}
+\frac{k_2^2-\viert}{\eta^2}\bigg)\right]\right\}\dt\right)
\nonumber\\   &&\!\!\!\!\!\!\!\!\!\!\!
         \\   &&\!\!\!\!\!\!\!\!\!\!\!
G^{(V_3)}(\xi'',\xi',\eta'',\eta';E)
=\int_0^\infty\d s''\e^{\i\alpha s''/\hbar}
K^{(V_3)}(\xi'',\xi',\eta'',\eta';s'')\enspace,
\end{eqnarray}
with the transformed path integral $K(s'')$ given by
\begin{eqnarray}
&&K^{(V_3)}(\xi'',\xi',\eta'',\eta';s'')
=\pathints{\xi}\exp\left[\ih\ints \left(\frac{m}{2}\dot\xi^2
+Eb\xi^2-\frac{\hbar^2}{2m}\frac{\lambda_1^2-\viert}{\xi^2}\right)\d s\right]
\nonumber\\   &&\qquad\times
\pathints{\eta}\exp\left[\ih\ints \left(\frac{m}{2}\dot\eta^2
+Eb\eta^2-\frac{\hbar^2}{2m}\frac{\lambda_2^2-\viert}{\eta^2}\right)\d s\right]
\nonumber\\   &&
K_{dis}^{(V_3)}(\xi'',\xi',\eta'',\eta';s'')=\sum_{n_\xi n_\eta=0}^\infty
\Psi_{n_\xi}^{(RHO,\lambda_1)}(\xi'')\Psi_{n_\xi}^{(RHO,\lambda_1)}(\xi')
\Psi_{n_\eta}^{(RHO,\lambda_2)}(\eta'')\Psi_{n_\eta}^{(RHO,\lambda_2)}(\eta')\,
\nonumber\\   &&\qquad\times
\exp\bigg[-\ih s''(2n_\xi+2n_\eta+\lambda_1+\lambda_2+2)\hbar\omega)\bigg]
\enspace.
\label{DIIV3dis-xi-eta}
\end{eqnarray}
We have inserted for the discrete spectrum the solution of the radial
harmonic oscillator in the usual way.
Performing the $s''$-integration gives the same spectrum as in
(\ref{EnlDIIV3}), as it should be.

The continuous spectrum is extracted in the usual way 
by means of (\ref{I-lambda-to-K}) and we obtain:
\begin{eqnarray}
&&G^{(V_3)}(\xi'',\xi',\eta'',\eta';E)
\nonumber\\   &&
=\sum_{n_\xi n_\eta=0}^\infty 
\frac{\hbar N_{ln}^2}{E_{ln}-E}
\Psi_{n_\xi}^{(RHO,\lambda_1)}(\xi'')\Psi_{n_\xi}^{(RHO,\lambda_1)}(\xi')
\Psi_{n_\eta}^{(RHO,\lambda_2)}(\eta'')
\Psi_{n_\eta}^{(RHO,\lambda_2)}(\eta')\,
\nonumber\\   &&
+\hbar(\xi'\xi''\eta'\eta'')^{-1/2}\int\d\CE\int_0^\infty
\frac{\d p\,p\sinh\pi p}{\frac{\hbar^2}{2m|a|}(p^2+k_<^2)-E}
\frac{|\Gamma[\half(1+\i p_1-\CE)]|^2|\Gamma[\half(1+\i p_2-\CE)]|^2}
{2\pi\tilde p^2}
\nonumber\\  &&\qquad\times
W_{\CE/2,\i p/2}\Big(\i\tilde p_1{\xi''}^2\Big)
W_{\CE/2,\i p/2}^*\Big(\i\tilde p_1{\xi'}^2\Big)
W_{\CE/2,\i p/2}\Big(\i\tilde p_2{\xi''}^2\Big)
W_{\CE/2,\i p/2}^*\Big(\i\tilde p_2{\eta'}^2\Big)
\end{eqnarray}
($\tilde p_{1,2}\equiv\sqrt{b(p^2+k_{1,2}^2)/|a|}$),
with the discrete energy spectrum as determined by (\ref{EnlDIIV3}), and
the normalization constant $N_{ln}$ resulting from the residuum in
(\ref{DIIV3dis-xi-eta}).

\subsection{The Superintegrable Potential $V_4$ on $\DII$.}
We consider the potential $V_4$ in the respective coordinate systems
\begin{eqnarray}
V_4(u,v)&=&\frac{u^2}{bu^2-a}\frac{\hbar^2}{2m}v_0^2
\\      &=&
\left(\frac{b\e^{2\tau_2}}{\cosh^2\tau_1}-a\right)^{-1}
\frac{\hbar^2v_0^2}{2m}\frac{\e^{2\tau_2}}{\cosh^2\tau_1}
\\      &=&
\left(\frac{b\xi^2\eta^2-a}{\xi^2\eta^2}\right)^{-1}
\frac{1}{\xi^2+\eta^2}\frac{\hbar^2v_0^2}{2m}(\xi^2+\eta^2)
\\      &=&
\left(\frac{bd^2\cosh^2\omega\cos^2\vphi-a}{\cosh^2\omega\cos^2\vphi}
\right)^{-1}
\frac{1}{\cosh^2\omega-\cos^2\vphi}\frac{\hbar^2v_0^2}{2m}
(\cosh^2\omega-\cos^2\vphi)\enspace.\qquad
\end{eqnarray}
We have displayed the potential in a somewhat more complicated way
to demonstrate the effect of the separation procedures.
The quantity $v_0$ enters the formulas in a way that only the
respective quantum numbers are altered. We will not go into the details,
and consider the potential $V_4$ only in the $(u,v)$-system. 
For the remaining systems we refer to \cite{GROas}.
Let us note that the separability of $V_4$ in all the four coordinate
systems on $\DII$ shows that a quantum system of a three-dimensional
analogue of $\DII$ is also separable in three-dimensional
``cylindrical'' versions of the $(u,v)$-system, spherical, parabolic
and elliptic coordinates \cite{GROat}. The additional quantum number associated
with the third coordinate can be identified with $v_0$.

Inserting $V_4$ into the path integral in the $(u,v)$-systems yields
\begin{eqnarray}
&&K^{(V_4)}(u'',u',v'',v';T)=\pathint{u}\pathint{v}\frac{bu^2-a}{u^2}
\nonumber\\ &&\qquad\times
\exp\left\{\ih\int_0^T\left[
\frac{m}{2}\frac{bu^2-a}{u^2}(\dot u^2+\dot v^2)
-\frac{u^2}{bu^2-a}\frac{\hbar^2v_0^2}{2m}\right]\dt\right\}
\nonumber\\  &&
=\int_{-\infty}^\infty\frac{\d E}{2\pi\hbar}\,\e^{-\i ET/\hbar}
   \int_0^\infty \d s''
\nonumber\\  &&\qquad\times
   \pathints{u}\pathints{v}
   \exp\left\{\ih\ints \bigg[
   \frac{m}{2}(\dot u^2+\dot v^2)-\frac{aE}{u^2}\bigg]\d s
    +\ih s''\bigg(bE-\frac{\hbar^2v_0^2}{2m}\bigg)\right\}
\nonumber\\  &&
=\int_{-\infty}^\infty\frac{\d E}{2\pi\hbar}\,\e^{-\i ET/\hbar}
\int_0^\infty\frac{ds''}{2\pi}\int_{-\infty}^\infty\d k\,\e^{\i k(v''-v')}
\exp\bigg(\ih bEs''-\ih\frac{\hbar^2}{2m}(k^2+v_0^2)s''\bigg)
\nonumber\\ &&\qquad\times
\pathints{u}\exp\left[\ih\ints \bigg(\frac{m}{2}\dot u^2
 -\hbar^2\frac{\lambda^2-\viert}{2mu^2}\bigg)\d s\right]
\nonumber\\  &&
=\int_{-\infty}^\infty\frac{\d E}{2\pi\hbar}\,\e^{-\i ET/\hbar}
\int_0^\infty ds''\int_{-\infty}^\infty\d k\,\e^{\i k(v''-v')}
\frac{m\sqrt{u'u''}}{\i\hbar s''}
\nonumber\\ &&\qquad\times
\exp\left[\ih\bigg(bE-\frac{\hbar^2}{2m}(k^2+v_0^2)\bigg)s''
  +\ih\frac{m}{2s''}({u'}^2+{u''}^2)\right]
  I_\lambda\bigg(\frac{mu'u''}{\i\hbar s''}\bigg)
\label{PathintegralDIIV4}
\end{eqnarray}
($\lambda^2-1/4=2maE/\hbar^2$).
We observe that the principal effect of the introduction of $V_4$
consists in a change in the quantum number $k$ which can be 
formulated as $\tilde k^2=k^2+v_0^2$. We can therefore write down the
solution by referring to \cite{GROas} and get
\begin{eqnarray}
&&G^{(V_4)}(u'',u',v'',v';E)
\nonumber\\  &&
=\frac{2m\sqrt{u'u''}}{\i\hbar}\int_{-\infty}^\infty\d k\,\e^{\i k(v''-v')}
I_\lambda\left(\sqrt{\tilde k^2-\frac{2mbE}{\hbar^2}}\,u_<\right)
K_\lambda\left(\sqrt{\tilde k^2-\frac{2mbE}{\hbar^2}}\,u_>\right)
\\  &&=
\frac{\hbar}{\pi^2}\int_{-\infty}^\infty\d k\,\frac{\e^{\i k(v''-v')}}{2\pi}
\nonumber\\  &&\qquad\times
\int_0^\infty\frac{2p\sinh\pi p\d p}{\frac{\hbar^2}{2m|a|}(p^2+\viert)-E}
K_{\i p}\left(\sqrt{\tilde k^2-\frac{2mbE}{\hbar^2}}\,u'\right)
K_{\i p}\left(\sqrt{\tilde k^2-\frac{2mbE}{\hbar^2}}\,u''\right)\enspace,\qquad
\label{GDarbouxIIuv}
\end{eqnarray}
with 
\begin{equation}
\lambda=\sqrt{\viert-\frac{2m|a|E}{\hbar^2}}\equiv \i p\enspace.
\end{equation}
The wave functions and the energy spectrum are read off:
\begin{eqnarray}
\Psi^{(V_4)}_{pk}(u,v)&=&\frac{\e^{\i k v}}{\sqrt{2\pi}}
\cdot\frac{\sqrt{2p\sinh\pi p}}{\pi}
K_{\i p}\left(\sqrt{\tilde k^2-\frac{2mbE}{\hbar^2}}\,u\right)\enspace,
\\
E&=&\frac{\hbar^2}{2m|a|}\bigg(p^2+\frac14\bigg)\enspace.
\end{eqnarray}


\begin{table}[t!]
\caption{\label{solutions} Solutions of the path integration for
  superintegrable potentials in   Darboux spaces}
\vspace{-0.5cm}
\hfuzz=20pt
\begin{eqnarray}
\hspace{2cm}
\begin{array}{l}\vbox{\small\offinterlineskip
\halign{&\vrule#&$\strut\ \hfil\hbox{#}\hfill\ $\cr
\noalign{\hrule}
height2pt&\omit&&\omit&\cr
&Space and Potential  &&Solution in terms of the wave-functions  &\cr
height2pt&\omit&&\omit&\cr
\noalign{\hrule}\noalign{\hrule}\noalign{\hrule}\noalign{\hrule}
height2pt&\omit&&\omit&\cr
&$\DI$                &&          &\cr
height2pt&\omit&&\omit&\cr
\noalign{\hrule}\noalign{\hrule}
height2pt&\omit&&\omit&\cr
&$V_1$:  $(u,v)$      &&Hermite polynomials $\times$ 
                        Parabolic cylinder functions      &\cr
&\qquad
Parabolic             &&No explicit solution              &\cr
height2pt&\omit&&\omit&\cr
\noalign{\hrule}
height2pt&\omit&&\omit&\cr
&$V_2$:  $(u,v)$      &&Hermite polynomials $\times$ 
                        Parabolic cylinder functions      &\cr
&\qquad
$(r,q)$               &&Hermite polynomials $\times$ 
                        Parabolic cylinder functions      &\cr
height2pt&\omit&&\omit&\cr
&$V_3$:  $(u,v)$      &&Product of Airy functions         &\cr
&\qquad
$(r,q)$               &&Product of Airy functions         &\cr
&\qquad
Parabolic             &&Product of Airy functions         &\cr
height2pt&\omit&&\omit&\cr
\noalign{\hrule}\noalign{\hrule}
height2pt&\omit&&\omit&\cr
&$\DII$               &&          &\cr
height2pt&\omit&&\omit&\cr
\noalign{\hrule}\noalign{\hrule}
height2pt&\omit&&\omit&\cr
&$V_1$:  $(u,v)$      &&Hermite polynomial $\times$
                        Whittaker functions$^*$           &\cr
&\qquad
Parabolic             &&No explicit solution              &\cr
height2pt&\omit&&\omit&\cr
\noalign{\hrule}
height2pt&\omit&&\omit&\cr
&$V_2$:  $(u,v)$      &&Laguerre polynomial $\times$
                        Whittaker functions$^*$           &\cr
&\qquad
Polar                 &&Gegenbauer polynomial $\times$
                        Whittaker functions$^*$           &\cr
&\qquad
Elliptic              &&No explicit solution              &\cr
height2pt&\omit&&\omit&\cr
\noalign{\hrule}
height2pt&\omit&&\omit&\cr
&$V_3$:  Polar        &&Gegenbauer polynomial $\times$
                        Whittaker functions$^*$           &\cr
&\qquad
Parabolic             &&Gegenbauer polynomial $\times$
                        Whittaker functions$^*$           &\cr
&\qquad
Displaced Elliptic    &&No explicit solution              &\cr
height2pt&\omit&&\omit&\cr
\noalign{\hrule}
height2pt&\omit&&\omit&\cr
&$V_4$:  $(u,v)$      &&Product of Bessel functions       &\cr
&\qquad
  Polar               &&Bessel functions $\times$
                        Legendre functions                &\cr
&\qquad
Parabolic             &&Product of Whittaker functions$^*$&\cr
&\qquad
Elliptic              &&Spheroidal wave-functions         &\cr
height2pt&\omit&&\omit&\cr
\noalign{\hrule}}}
\nonumber\\
\hspace{-2cm}
\hbox{($^*$: The notion Whittaker functions means in all cases for a 
discrete spectrum Laguerre poly-}
\nonumber\\
\hspace{-2cm}
\hbox{nomials, and for a continuous spectrum Whittaker
functions $W_{\mu,\nu}(z)$, respectievly $M_{\mu,\nu}(z)$.)}
\end{array}
\nonumber\end{eqnarray}
\vspace{-0.5cm}
\end{table}
\hfuzz=5pt

\setcounter{equation}{0}%
\section{Summary and Discussion}
\message{Summary and Discussion}
In this paper we have discussed superintegrable potentials on spaces
of non-constant curvature. The results are very satisfactory.
According to \cite{DASYPS,KalninsKMWinter,KalninsKWinter} there are three
potentials on $\DI$, four potentials on $\DII$, five potentials on
$\DIII$, and four potentials on $\DIV$, respectively.
We could solve many of the emerging quantum mechanical problems.
To give an overview, we summarize our results in Table
\ref{solutions}. We list for each space the corresponding potentials
including the general form of the solution (if explicitly
possible). We omit the trivial potentials here, because they are
separable in all corresponding coordinate systems.

We were able to solve the various path integral representations, because we
have now to our disposal not only the basic path integrals for the harmonic 
oscillator, the linear oscillator, the radial harmonic oscillator, and
the P\"oschl--Teller Potential, but also path integral
identities derived from path integration on harmonic spaces like the
elliptic and spheroidal path integral representations with its more 
complicated special functions~\cite{GROad,GKPSa,GRSh}. 
This includes also numerous transformation
techniques to find a particular solution based on one of the basic solutions.
Various analysis techniques can be applied to find not
only an expression for the Green  function but also for the
wave-functions and the energy spectrum. 

We also observe a new feature of superintegrable potentials. We learned from
our investigation of potential problems on $\DI$ that degeneracy
for superintegrable potentials does not follow automatically.
In fact, our (counter-)examples show that the usually accepted
opinion that superintegrability and degeneracy of a quantum system 
are equivalent statements is not true in general. It would be interesting to
formulate the precise additional mathematical requirements that these
statements are actually true in general. In our case the non-equivalence
of these two notions comes from the boundary-conditions which had to imposed
on $\DI$ in order to guarantee a well-defined Hilbert space.

We found in all cases a discrete and a continuous spectrum for the 
superintegrable potentials. We also could compare some limiting cases,
e.g. for the Darboux space $\DII$, where we could recover the
corresponding solutions for the two-dimensional Euclidean space and
the two-dimensional hyperboloid. On $\DI$ the energy spectra are only
determined by a transcendental equation due to the boundary condition
for the coordinate $u$.
On $\DII$ we found analogues of the singular oscillator, the Holt potential
and the Coulomb potential in two dimensional Euclidean space. We could recover
these limiting cases in the equations for the energy spectra.
The equations equations for the energy spectra were on $\DII$ algebraic
equations in second and fourth order in the energy. This allows several
solutions depending on the specific values of the parameters $a$ and $b$ and
possible further boundary conditions. Also semi-bound states may be possible.

In a forthcoming publications we will treat the two remaining Darboux spaces
$\DIII$ and $\DIV$, respectively. In particular on $\DIII$ there is already a 
discrete spectrum possible for the free motion, and has the form
\begin{equation}
E_{nl}=-\frac{\hbar^2}{2m}\frac{b}{a^2}(2n+2l+1)^2\enspace.
\end{equation}
yielding for $b>0$ an infinite number of bound states.
This is similar as the motion on the $\SU(1,1)$ hyperboloid, where a
continuous and a discrete spectrum exists \cite{BJb}.
On $\DIII$ there are five superintegrable potentials and on $\DIV$ there are 
four superintegrable potentials.

Let us finally discuss the following issue: Let us consider a
three-dimensional generalization of the Darboux space $\DII$
with a line element
\begin{equation}
\d s^2=\frac{bu^2-a}{u^2}(\d u^2+\d v^2+\d w^2)\enspace,
\label{DarbouxII-3D}
\end{equation} 
and $w$ is the new variable.
$\DII$ has the property that for $a=0$, $b=1$, we recover the
two-dimensional Euclidean plane, and all four coordinate systems on
the two-dimensional Euclidean plane are also separable  coordinate
systems on $\DII$ for the Schr\"odinger, respectively the Helmholtz equation.
However, in order to set up a well-defined quantum theory a curvature term 
$(\hbar^2/2m)\cdot(R/8)$ must be introduced in the quantum Hamiltonian
\cite{GROat,KaMih}.
In the present case of (\ref{DarbouxII-3D}), which we might call
three-dimensional Darboux space II, for short $\3dDII$, it is easily
checked that all eleven systems for the three-dimensional Euclidean plane
which separate the Schr\"odinger, respectively the Helmholtz equation,
also separate the Schr\"odinger, respectively the Helmholtz equation
on $\3dDII$. As it has been shown in \cite{GROat}, the corresponding quantum
motion can be explicitly evaluated by means of the path integral with energy
spectrum
\begin{equation}
E=\frac{\hbar^2}{2m|a|}(p^2+1)\enspace.
\end{equation}
As it is well known, there are four minimally superintegrable
potentials in three-dimensional Euclidean space and five maximally
superintegrable potentials, and it is obvious how to construct 
maximally superintegrable potentials on $\3dDII$.
In a forthcoming publication the details will be worked out.

\subsection*{\bf Acknowledgments}
This work was supported by the Heisenberg--Landau program. 

The authors are grateful to Ernie Kalnins for fruitful and pleasant
discussions on superintegrability and separating coordinate systems. 
C.Grosche would like to thank the organizers of the XII. International
Conference on Symmetry Methods in Physics, July 3--8, Yerevan, Armenia,
for the warm hospitality during the stay in Yerevan.

G.S.Pogosyan acknowledges the support of the Direcci\'on General de Asuntos
del Personal Acad\'emico, Universidad Nacional Aut\'onoma de M\'exico 
({\sc dgapa--unam}) by the grant 102603 {\it Optica Matem\'atica}, 
{\sc sep-conacyt} project 44845 and {\sl PROMEP} 103.5/05/1705.


\input cyracc.def
\font\tencyr=wncyr10
\font\tenitcyr=wncyi10
\font\tencpcyr=wncysc10
\def\cyrrm{\tencyr\cyracc}
\def\cyrit{\tenitcyr\cyracc}
\def\cyrcp{\tencpcyr\cyracc}
\newpage\noindent%
\renewcommand{\baselinestretch}{0.9}%
\noindent%

\bigskip
\vbox{\centerline{\ }
\centerline{\quad\epsfig{file=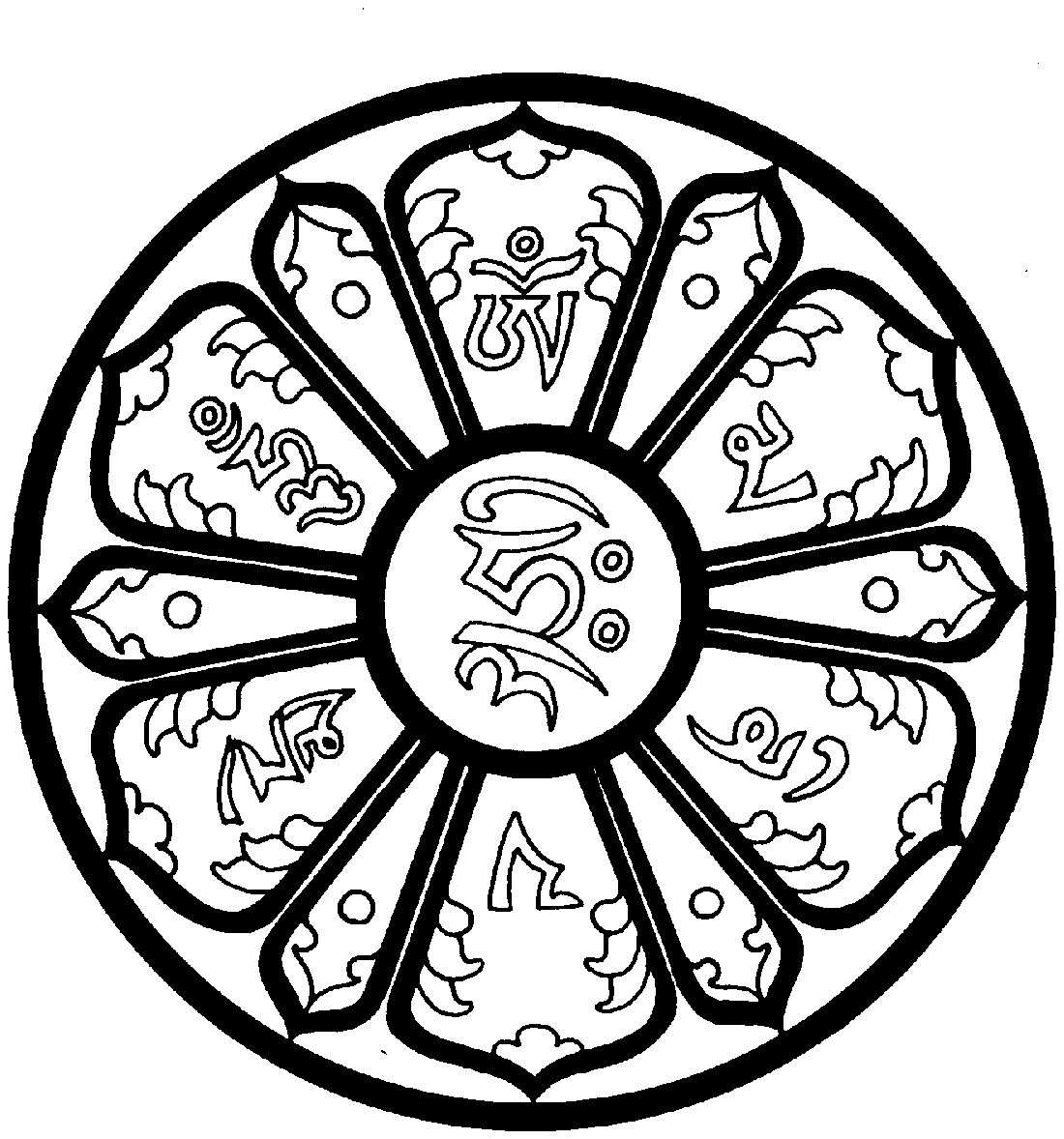,width=4cm}}}

\end{document}

%% file: posfig.tex
\makeatletter
\setlength\@fptop{0\p@ }
\setlength\@fpsep{12\p@ }
\setlength\@fpbot{0\p@ plus 1fil }
\makeatother

\def\textfraction{.01}
\def\floatpagefraction{.8}

\intextsep 20pt plus 2pt minus 2pt
\setcounter{topnumber}{4}
\def\topfraction{.9}
\setcounter{bottomnumber}{2}
\def\bottomfraction{.8}
\setcounter{totalnumber}{6}

%
\makeatletter
\newinsert \@kludgeins
\global\dimen\@kludgeins \maxdimen
\global\count\@kludgeins 1000
\gdef \enlargethispage {%
   \@ifstar
     {%
      \@enlargepage{\hbox{\kern\p@}}}%
     {%
      \@enlargepage\@empty}%
}
\gdef\@enlargepage#1#2{%
   \@tempskipa#2\relax
   \ifdim \@tempskipa>.5\maxdimen
     \@latexerr{Suggested\space extra\space height\space
                (\the\@tempskipa)\space dangerously\space
                large}\@eha
   \else
     \ifdim \vsize<.5\maxdimen
       \@bsphack
         \insert\@kludgeins{#1\vskip-\@tempskipa}%
       \@esphack
     \else
       \@latexerr{Page\space height\space already\space
                  too\space large}\@eha
     \fi
   \fi
}
\makeatother